\documentclass[11pt,a4paper]{article}

\usepackage{latexsym}
\usepackage{amsmath}
\usepackage{amssymb}
\usepackage{amsthm}
\usepackage{amscd}
\usepackage[dvips]{graphicx}

\usepackage{enumerate}

\theoremstyle{plain}    
\newtheorem{thm}{Theorem}[section]
\numberwithin{equation}{section} 
\numberwithin{figure}{section} 
\theoremstyle{plain}    
\newtheorem{cor}[thm]{Corollary} 
\theoremstyle{plain}    
\newtheorem{lem}[thm]{Lemma} 
\theoremstyle{plain}    
\newtheorem{prop}[thm]{Proposition} 
\theoremstyle{definition}
\newtheorem{defn}[thm]{Definition}

\makeatletter
 
 \@addtoreset{equation}{section}
\makeatother

\renewcommand{\labelenumi}{\rm(\theenumi)}

\def\rnum#1{\expandafter{\romannumeral #1}}
\def\Rnum#1{\uppercase\expandafter{\romannum}}

\newcommand{\Lap}{\mathit{\Delta}}
\makeatother

\begin{document}

\newpage \thispagestyle{empty}
{\topskip 2cm
\begin{center}
 \textbf{\huge On Irreducibility of the Energy Representation of the Gauge
Group and the White Noise Distribution Theory}{\huge \par}

\bigskip
\bigskip
\bigskip
\bigskip
\bigskip
\bigskip

{\large Yoshihito Shimada}
\bigskip 

\textit{Graduate School of Mathematics}

\textit{Kyushu University}

\textit{1-10-6 Hakozaki, Fukuoka 812-8581}

\textit{JAPAN}
\end{center}
\vfil
\noindent
We consider the energy representation for the gauge group.
The gauge group is the set of \( C^{\infty } \)-mappings from 
a compact Riemannian manifold to a semi-simple compact Lie group.  
In this paper, we obtain irreducibility
of the energy representation of the gauge group 
for any dimension of \( M \).
To prove irreducibility for the energy representation, we use the fact that
each operator from a space of test functionals to a space of generalized
functionals is realized as a series of integral kernel operators, called 
the Fock expansion.

\noindent
\bigskip
\hrule

\noindent
\texttt{\textbf{\small e-mail:shimada@math.kyushu-u.ac.jp}}
\vfil
\newpage
\setcounter{page}{1}

\section{Introduction}

In this paper, we discuss irreducibility of the energy representation of the
gauge group. The gauge group is the set of all \( C^{\infty } \)-mappings with
compact support from a Riemannian manifold \( M \) to a semi-simple compact
Lie group \( G \). Then the energy representation of the gauge group is studied
in \cite{Albe-paper}, \cite{Albe_1-dim_case}, \cite{Gelfand_et_al(1977)}, 
\cite{Gelfand_et_al}, and \cite{Ismagilov-paper}.
The first definition of the energy representation appears 
in \cite{Versik_et_al} and  
I. Gelfand et al. proved irreducibility in case of 
simple compact Lie group \( G\) and \(\dim M\geq 2\) 
in \cite{Gelfand_et_al(1977)}. 
(Unfortunately, the proof of irreducibility 
in \cite{Gelfand_et_al(1977)} has a gap. 
However they succeeded in proving irreducibility 
for the case \(\dim M\geq 4\) in \cite{Gelfand_et_al}.)
R. Ismagilov \cite{Ismagilov-paper} showed irreducibility of 
the energy representation for the case \( \dim M\geq 5 \) 
and \( G=SU(2) \).
S. Albeverio et al.,in \cite{Albe-paper}, proved for 
the case \( \dim M\geq 3 \) and obtained the partial solution 
for the same question for the case \( \dim M=2 \).
In case of \( \dim M=1 \), S. Albeverio et al., in \cite{Albe-paper} and
\cite{Albe_1-dim_case}, showed reducibility for the energy representation of
a Sobolev-Lie group \( H(\mathbf{R},G) \) and of a subgroup of \( H(S^{1},G) \)
consisting of all based loops. In section 4 of this paper, 
we shall obtain irreducibility of the energy representation 
of \( C^{\infty }(M,G) \) for any dimension of
\( M \) when \( M \) is compact. 

We present a brief sketch of our idea for proof of irreducibility.
In order to prove irreducibility for the case \( \dim M\geq 2 \), 
all authors of previous works used results of 
a Gaussian measure \( \mu  \) on the real vector space \( E^{*} \) of
distributions. On the other hand, in order to analyze 
operators on \( L^{2}(E^{*},\mu ) \), we can use the theory
of the spaces \( (E) \) and \( (E)^{*} \), and of operators from \( (E) \)
to \( (E)^{*} \), called the white noise calculus. 
Here \( (E) \) is a space of test functionals and \( (E)^{*} \) is a space
of generalized functionals.
The white noise calculus is introduced by T. Hida in 1975, 
and led us to important consequences concerning
to an analysis of the Boson system. (See \cite{Obata}.) 
In the white noise calculus, a continuous linear operators from \( (E) \)
to \( (E)^{*} \) is realized as a series of operators \( \Xi _{l,m}(\kappa _{l,m}) \),
\( l,m\in \mathbf{Z}_{\geq 0} \) such that
\[
  \Xi _{l,m}(\kappa _{l,m})
    =\int \kappa _{l,m}(s_{1},\ldots ,s_{l},t_{1},\ldots ,t_{m})
      \partial ^{*}_{s_{1}}\ldots \partial ^{*}_{s_{l}}
      \partial _{t_{1}}\ldots \partial _{t_{m}}ds_{1}
      \ldots ds_{l}dt_{1}\ldots dt_{m}
\]
where \( \partial ^{*}_{s} \) is a creation operator, \( \partial _{t} \)
is an annihilation operator, and \( \kappa _{l,m} \) is a kernel distribution.
\( \Xi _{l,m}(\kappa _{l,m}) \) is called an integral kernel operator with a
kernel distribution \( \kappa _{l,m} \) and this realization is called Fock
expansion for a continuous linear operator from \( (E) \) to \( (E)^{*} \).

The merit of using the Fock expansion is that 
we can determine the commutant by direct algebraic computation, 
that is, the problem of irreducibility becomes easy comparatively. 
It is not exaggeration to say like this since we have the following reason. 
We know that one possible way to determine the commutant 
is to apply Tomita-Takesaki modular theory of the von Neumann algebras.
However, the existence of the cyclic and separating vector 
for von Neumann algebra generated by the energy representation
is not obvious. In fact, in \cite{Albe-paper} and \cite{Albe_1-dim_case}, 
S. Albeverio et al. succeeded in proving reducibility 
for the energy representation of a subgroup of \( H(S^{1},G) \) 
only in the cyclic component with respect to the vacuum vector.
However, they failed to show that vacuum vector is cyclic.

We should remark the relation between our result and the result of previous
works \cite{Albe-paper} and \cite{Albe_1-dim_case}. 
Albeverio et al. \cite{Albe-paper} considered the energy representation 
of the Sobolev-Lie group \( H(S^{1},G) \). 
They showed that the cyclic component of the vacuum of
the energy representation of \( H(S^{1},G) \) is unitary equivalent to
\begin{gather*}
  (U^{R}(\phi )f)(\eta )
    :=\left( \frac{d\mu (\eta \phi )}{d\mu (\eta )}\right) ^{\frac{1}{2}}
        f(\eta \phi ), \\
  \phi \in H(S^{1},G), \quad 
  f\in L^{2}(\{\eta \in C(S^{1},G)\, |\, \eta (0)=\eta (2\pi )=e\};\mu),
\end{gather*}
and showed reducibility of \( U^{R} \). 
Here \( H(S^{1},G) \) is the completion of all smooth loops \(\phi \)
satisfying \(\phi (0)=\phi (2\pi )=e \).
On the other hand, our result is related with the group
\( C^{\infty }(S^{1},G) \), where \( \phi \in C^{\infty }(S^{1},G) \)
needs not satisfy the condition \( \phi (0)=\phi (2\pi )=e \). 
Therefore, our result on irreducibility is not in conflict with 
reducibility obtained in \cite{Albe-paper} and \cite{Albe_1-dim_case}. 
The difference of two above-mentioned results on irreducibility
is also seen in the work of V. Jones and A. Wassermann. 
It is well-known fact that 
the level \( \ell \) projective unitary representation
\( \pi ^{\otimes \ell } \) of \( C^{\infty }(S^{1},G) \) is irreducible.
However A. Wassermann proved that the von Neumann algebra
\[
  \pi ^{\otimes \ell }(\{\phi \in C^{\infty }(S^{1},G)\, 
    |\, \phi (\theta )=e\, \, 
    \mathrm {for\, all\, }\theta \in [\pi ,2\pi ]\})''
\]
is type III\(_1\) factor (Theorem A of \cite{Wassermann}).

Now we present the organization of this paper. In section 2, we briefly sketch
the white noise calculus and prepare for our analysis of the energy representation.
In section 3, we define the gauge group and its energy representation. In section
4, the main theorem \ref{thm:main-theorem} is mentioned and proved.

\section{Survey of the white noise calculus}\label{section:white-noise}

In this section, we introduce the white noise calculus. The details of this
section are in \cite{Obata}.

\begin{defn}
\label{thm:property_of_self-adj_op_A}Let \( H \) be a complex Hilbert space
with an inner product \( \left\langle \cdot ,\cdot \right\rangle _{0} \). Let
\( A \) be a self-adjoint operator defined on a dense domain \( D(A) \). Let
\( \{\lambda _{j}\}_{j\in \mathbf{N}} \) be eigenvalues of \( A \)
and \( \{e_{j}\}_{j\in \mathbf{N}} \) be normalized eigenvectors
for \( \{\lambda _{j}\}_{j\in \mathbf{N}} \), i.e., \( Ae_{j}=\lambda _{j}e_{j} \)
and \( |e_{j}|_{0}=1 \) for all \( j\in \mathbf{N} \). 
Moreover, we also assume the following two conditions:
\renewcommand{\theenumi}{\roman{enumi}}
\renewcommand{\labelenumi}{\rm(\theenumi)}
\begin{enumerate}
\item \( \{e_{j}\}_{j\in \mathbf{N}} \) is a C.O.N.S. of \( H \), 
\item Multiplicity of \( \{\lambda _{j}\}_{j\in \mathbf{N}} \) is finite and
\( 1<\lambda _{1}\leq \lambda _{2}\leq \ldots \rightarrow \infty  \). 
\end{enumerate}
\renewcommand{\theenumi}{\arabic{enumi}}
\renewcommand{\labelenumi}{\rm(\theenumi)}
\noindent Then we have the following properties.

\begin{enumerate}
\item For \( p\in \mathbf{Z}_{\geq 0} \) and \( x,\, y\in D(A^{p}) \), let \( \left\langle x,y\right\rangle _{p}:=\left\langle A^{p}x,A^{p}y\right\rangle _{0} \).
Then \( \left\langle \cdot ,\cdot \right\rangle _{p} \) is an inner product
on \( D(A^{p}) \). Moreover, \( D(A^{p}) \) is complete with respect to the
norm \( |\cdot |_{p} \), that is, the pair
\( E_{p}:=(D(A^{p}),\left\langle \cdot ,\cdot \right\rangle _{p}) \)
is a Hilbert space. 
\item For \( p\geq 0 \), let \( j_{p,p+1}:E_{p+1}\rightarrow E_{p} \) be the inclusion
map. Then every inclusion map is continuous and has a dense image. For \( q\geq p\geq 0 \),
let 
\[
j_{p,q}:=j_{p,p+1}\circ \ldots \circ j_{q-1,q}:E_{q}\rightarrow E_{p}.\]
Then \( \{E_{p},j_{p,q}\} \) is a reduced projective system. 
\item A {\it countable Hilbert space}
\[
E:=\lim _{\leftarrow }E_{p}=\bigcap _{p\geq 0}E_{p}\]
constructed from the pair \( (H,A) \) is a reflexive Fr{\'e}chet space.
We call \( E \) {\it CH-space} simply.
\item From (3), we have \( \displaystyle E^{*}=\lim _{\rightarrow }E^{*}_{p} \) as
a topological vector space, i.e. the strong topology on \( E^{*} \) and the
inductive topology on \( \displaystyle \lim _{\rightarrow }E^{*}_{p} \) coincide.
\item Let \( p\in \mathbf{Z}_{\geq 0} \) and \( \left\langle x,y\right\rangle _{-p}:=\left\langle A^{-p}x,A^{-p}y\right\rangle _{0} \).
Then \( \left\langle \cdot ,\cdot \right\rangle _{-p} \) is an inner product
on \( H \). 
\item For \( p\geq 0 \), let \( E_{-p} \) be the completion of \( H \) with respect
to the norm \( |\cdot |_{-p} \). Then we can consider the inclusion map \( i_{-(p+1),-p}:E_{-p}\rightarrow E_{-(p+1)} \),
and for \( q\geq p\geq 0 \) let
\[
i_{-q,-p}:=i_{-q,-q+1}\circ \ldots \circ i_{-(p+1),-p}:E_{-p}\rightarrow E_{-q}.\]
 Then \( \{E_{-p},i_{-q,-p}\} \) is an inductive system. Moreover, \( E_{-p} \)
and \( E_{p}^{*} \) are anti-linear isomorphic and isometric. Thus, from (4),
we have
\[
E^{*}=\lim _{\rightarrow }E_{-p}=\bigcup _{p\geq 0}E_{-p}.\]
 
\end{enumerate}
\end{defn}

Furthermore, we require for the operator \( A \) that there exists \( \alpha >0 \)
such that \( A^{-\alpha } \) is a Hilbert-Schmidt class operator, namely
\begin{equation}
  \delta ^{2}:=\sum _{j=1}^{\infty }
     \lambda _{j}^{-2\alpha }<\infty .\label{eq:Hilbert-Schmidt_inverse}
\end{equation}
From this condition, \( E \) (resp. \( E^{*} \)) is a nuclear space. Thus
we can define the \( \pi  \)-tensor topology \( E\otimes _{\pi }E \) (resp.
\( E^{*}\otimes _{\pi }E^{*} \)) of \( E \) (resp. \( E^{*} \)). If there
is no danger of confusion, we will use the notation \( E\otimes E \) (resp.
\( E^{*}\otimes E^{*} \)) simply.

We denote the canonical bilinear form on \( E^{*}\times E \) by \( \left\langle \cdot ,\cdot \right\rangle  \).
We have the following natural relation between the canonical bilinear form on
\( E^{*}\times E \) and the inner product on \( H \) :
\[
\left\langle f,g\right\rangle =\left\langle \overline{f},g\right\rangle _{0}\]
for all \( f\in H \) and \( g\in E \).

\begin{defn}
\label{thm:definition_of_symmetrizer}Let \( X \) be a Hilbert space, or a CH-space.
\begin{enumerate}
\item Let \( g_{1} \), \( \ldots  \) , \( g_{n}\in X \). We define the symmetrization
\( s_{n}(g_{1}\otimes \ldots \otimes g_{n}) \) of 
\( g_{1}\otimes \ldots \otimes g_{n}\in X^{\otimes n} \)
as follows.
\[
s_{n}(g_{1}\otimes \ldots \otimes g_{n}):=g_{1}\widehat{\otimes }
\ldots \widehat{\otimes }g_{n}:=\frac{1}{n!}\sum _{\sigma \in 
\mathfrak {S}_{n}}g_{\sigma (1)}\otimes \ldots \otimes g_{\sigma (n)},\]
where \( \mathfrak {S}_{n} \) is the set of all permutations of \( \{1,2,\ldots ,n\} \). 
\item If \( f\in X^{\otimes n} \) satisfies \( s_{n}(f)=f \), then we call \( f \)
symmetric. We denote the set of all symmetric elements of \( X^{\otimes n} \)
by \( X^{\widehat{\otimes }n} \) and we call \( X^{\widehat{\otimes }n} \)
the \( n \)-th symmetric tensor of \( X \). Then \( s_{n} \) is a projection
from \( X^{\otimes n} \) to \( X^{\widehat{\otimes }n} \) .
\item For \( F\in (X^{\otimes n})^{*} \) and \( \sigma \in \mathfrak {S}_{n} \),
let \( F^{\sigma } \) be an element of \( (X^{\otimes n})^{*} \) satisfying
\[
\left\langle F^{\sigma },g_{1}\otimes \ldots \otimes g_{n}\right\rangle :=\left\langle F,g_{\sigma ^{-1}(1)}\otimes \ldots \otimes g_{\sigma ^{-1}(n)}\right\rangle ,\quad g_{i}\in X.\]
Then we define the symmetrization \( s_{n}(F) \) as follows.
\[
s_{n}(F):=\frac{1}{n!}\sum _{\sigma \in \mathfrak {S}_{n}}F^{\sigma }.\]

\item If \( F\in (X^{\otimes n})^{*} \) satisfies \( s_{n}(F)=F \), we call \( F \)
symmetric. We denote the set of all symmetric elements of \( (X^{\otimes n})^{*} \)
by \( (X^{\otimes n})^{*}_{\mathrm{sym}} \).
\end{enumerate}
\end{defn}

From the above discussion, we obtain a Gelfand triple :
\[
E\subset H\subset E^{*}.\]

Next, we define the Fock space and the second quantization of a linear operator.

\begin{defn}
Let \( H \) be a Hilbert space and \( A \) be a linear operator on \( H \).

\begin{enumerate}
\item Let
\begin{gather*}
\Gamma _{\mathrm{b}}(H):=\left\{ \sum _{n=0}^{\infty }f_{n}\, 
        \Big|\, f_{n}\in H^{\widehat{\otimes }n},\, \sum _{n=0}^{\infty }
        n! \left| f_{n}\right| ^2_{0}<+\infty \right\} ,\\
\left\langle \!\!\! \left\langle \sum _{n=0}^{\infty }f_{n},
    \sum _{n=0}^{\infty }g_{n}\right\rangle \!\!\! \right\rangle _{\! \! \! 0}
    :=\sum _{n\in \mathbf{Z}_{\geq 0}}n!\left\langle f_{n},g_{n}\right\rangle _{0}.
\end{gather*}
Then we call \( \Gamma _{\mathrm{b}}(H) \) the {\it Boson Fock space}. The Boson
Fock space \( \Gamma _{\mathrm{b}}(H) \) is a Hilbert space with respect to
the inner product \( \left\langle \! \left\langle \cdot ,
\cdot \right\rangle \! \right\rangle _{0} \). 

\item Let ``\( \mathrm{id} \)'' be the identity operator on \( H \) and
\( \mathrm{id}_{m}:=\overbrace{\mathrm{id}\otimes \ldots \otimes \mathrm{id}}^{m} \). Let
\begin{gather*}
\Gamma _{\mathrm{b}}(A):=\sum _{n=0}^{\infty }A^{\otimes n},\\
d\Gamma _{\mathrm{b}}(A)^{(n)}
    :=\sum _{j=1}^{n}\mathrm{id}_{j-1}\otimes A\otimes \mathrm{id}_{n-j}, \quad and\\
d\Gamma _{\mathrm{b}}(A):=\sum _{n=0}^{\infty }d\Gamma _{\mathrm{b}}(A)^{(n)} .
\end{gather*}
Then we
call \( \Gamma _{\mathrm{b}}(A) \) the {\it second quantization} of \( A \) and
\( d\Gamma _{\mathrm{b}}(A) \) the {\it differential second quantization} of \( A \).
\end{enumerate}
\end{defn}

\begin{defn}
Let \( H \) be a complex Hilbert space and \( A \) be a self-adjoint operator
on \( H \) satisfying the conditions (i) and (ii) in definition \ref{thm:property_of_self-adj_op_A}
and \eqref{eq:Hilbert-Schmidt_inverse}. Then we can define a CH-space \( (E) \)
constructed from \( (\Gamma _{\mathrm{b}}(H),\Gamma _{\mathrm{b}}(A)) \) and
we obtain a Gelfand triple :
\[
(E)\subset \Gamma _{\mathrm{b}}(H)\subset (E)^{*}.\]
\end{defn}

\begin{cor}
Let \( \displaystyle \phi :=\sum _{n=0}^{\infty }f_{n}\in \Gamma _{\mathrm{b}}(H) \), \( f_{n}\in H^{\widehat{\otimes }n} \).
Then \( \phi \in (E) \) if and only if \( f_{n}\in E^{\widehat{\otimes }n} \)
for all \( n\geq 0 \). Moreover, it follows that
\[
\left\Vert \phi \right\Vert _{p}:=\left\Vert \Gamma _{\mathrm{b}}(A)^{p}\phi \right\Vert _{0}<+\infty \]
for all \( p\geq 0 \).
\end{cor}
We refer to a continuous linear operator on a locally convex space.

\begin{defn}
Let \( X \), \( Y \) be locally convex spaces. \( \mathcal{L}(X,Y) \) is
the set of all continuous linear operators from \( X \) to \( Y \). 
\end{defn}

\begin{lem}
Let \( X \) {\rm (}resp. \( Y \){\rm )} be a locally convex space with seminorms
\( \{|\cdot |_{X,q}\}_{q\in Q} \) {\rm (}resp. \( \{|\cdot |_{Y,p}\}_{p\in P} \){\rm )}.
Then a linear operator \( V:X\rightarrow Y \) is continuous, namely, 
\( V\in \mathcal{L}(X,Y) \) if and only if for any \( p\in P \), 
there exist \( q\in Q \) and \( C>0 \) such that
\[
|Vf|_{Y,p}\leq C|f|_{X,q},\quad f\in X.\]

\end{lem}
In order to discuss an integral kernel operator, we define a contraction of
tensor products.

\begin{defn}
Let \( H \) be a complex Hilbert space and \( A \) be a self-adjoint operator
on \( H \) satisfying the conditions (i) and (ii) in definition \ref{thm:property_of_self-adj_op_A}
and \eqref{eq:Hilbert-Schmidt_inverse}. 
Let
\[
e(\mathbf{i}):=e_{i_{1}}\otimes \ldots \otimes e_{i_{l}},\quad
\mathbf{i}:=(i_{1},\ldots ,i_{l})\in \mathbf{N}^{l}.\]

\begin{enumerate}
\item For \( F\in \left( E^{\otimes (l+m)}\right) ^{*} \), let
\[
|F|_{l,m;p,q}^{2}:=\sum _{\mathbf{i},\mathbf{j}}
   \left| \left\langle F,e(\mathbf{i})\otimes e(\mathbf{j})\right\rangle \right| ^{2}
   \left| e(\mathbf{i})\right| ^{2}_{p}\left| e(\mathbf{j})\right| ^{2}_{q}
\]
where \( \mathbf{i} \) and \( \mathbf{j} \) run over the whole of \( \mathbf{N}^{l} \)
and \( \mathbf{N}^{m} \) respectively.

\item For \( F\in \left( E^{\otimes (l+m)}\right) ^{*} \) and
\( g\in E^{\otimes (l+n)} \),
we define a contraction \( F\otimes _{l}g\in \left( E^{m+n}\right) ^{*} \)
of \( F \) and \( g \) as follows.
\[
F\otimes _{l}g:=\sum _{\mathbf{j},\mathbf{k}}
    \left( \sum _{\mathbf{i}}\left\langle F,e(\mathbf{j})\otimes e(\mathbf{i})\right\rangle
    \left\langle g,e(\mathbf{k})\otimes e(\mathbf{i})\right\rangle \right)
    e(\mathbf{j})\otimes e(\mathbf{k})
\]
where \( \mathbf{i} \), \( \mathbf{j} \), and \( \mathbf{k} \) run over the whole of 
\( \mathbf{N}^{l} \), \( \mathbf{N}^{m} \), and \( \mathbf{N}^{n} \) respectively. 
\end{enumerate}
\end{defn}

We check well-definedness of the contraction. For any \( F\in \left( E^{\otimes (l+m)}\right) ^{*} \)
there exists \( p\geq 0 \) such that \( |F|_{l,m;-p,-p}=|F|_{-p}<+\infty  \).
We note that \(|e(\mathbf{i})|_{-p}|e(\mathbf{i})|_{p}=1\)
for all \( p\geq 0 \) and
\( |e(\mathbf{i})|_{p}\leq |e(\mathbf{i})|_{p+q}\)
for \( p\in \mathbf{Z} \) and \( q\geq 0 \). Then we have
\[
\begin{split}
 |&F \otimes _{l}g|^{2}_{-p} \\
 &= |F\otimes _{l}g|^{2}_{l,m;-p,-p} \\
 &= \sum _{\mathbf{j},\mathbf{k}}
    \left| \sum _{\mathbf{i}}\left\langle F,e(\mathbf{j})\otimes e(\mathbf{i})\right\rangle
    \left\langle g,e(\mathbf{k})\otimes e(\mathbf{i})\right\rangle
    |e(\mathbf{i})|_{-p}|e(\mathbf{i})|_{p}\right| ^{2}
    \left| e(\mathbf{j})\right| _{-p}^{2}\left| e(\mathbf{k})\right| _{-p}^{2}\\
 &\leq  \sum _{\mathbf{j},\mathbf{k}}
   \left( \sum _{\mathbf{i}}\left| \left\langle F,e(\mathbf{j})
      \otimes e(\mathbf{i})\right\rangle \right| ^{2}|e(\mathbf{i})|^{2}_{-p}\right)
   \left( \sum _{\mathbf{i}'}\left| \left\langle g,e(\mathbf{k})
      \otimes e(\mathbf{i}')\right\rangle \right| ^{2}|e(\mathbf{i}')|^{2}_{p}\right)
   \left| e(\mathbf{j})\right| _{-p}^{2}\left| e(\mathbf{k})\right| _{-p}^{2}\\
 &\leq |F|_{-p}|g|_{p}.
\end{split}
\]
Therefore we obtain \( F\otimes _{l}g\in \left( E^{\otimes (l+m)}\right) ^{*} \).

Now we define an integral kernel operator.

\begin{defn}
[\bf Integral kernel operator] Let \( \kappa \in (E^{\otimes (l+m)})^{*} \).
For \( \phi :=\sum _{n=0}^{\infty }f_{n}\in (E) \), \( f_{n}\in E^{\widehat{\otimes }n} \),
let
\[
  \Xi _{l,m}(\kappa )\phi :=\sum _{n=0}^{\infty }\frac{(n+m)!}{n!}
    s_{l+m} \left( \kappa \otimes _{m}f_{m+n} \right) .
\]
Then \( \Xi _{l,m}(\kappa )\in \mathcal{L}((E),(E)^{*}) \). 
We call \( \Xi _{l,m}(\kappa ) \) an {\it integral kernel operator}
with a kernel distribution \( \kappa  \). 
\end{defn}
As for integral kernel operators, see section 4.3 of \cite{Obata}.
Note that the following map
\[
(E^{\otimes (l+m)})^{*}\ni \kappa \mapsto \Xi _{l,m}(\kappa )\in \mathcal{L}((E),(E)^{*})\]
is not injective. We define
\[
s_{l,m}(\kappa ):=\frac{1}{l!m!}\sum _{\sigma \in \mathfrak {S}_{l}\times \mathfrak {S}_{m}}\kappa ^{\sigma },\]
where \( \kappa ^{\sigma } \) is defined in definition \ref{thm:definition_of_symmetrizer}
(3). Put
\[
(E^{\otimes (l+m)})^{*}_{\mathrm{sym}(l,m)}:=\{\kappa \in (E^{\otimes (l+m)})^{*}\, |\, s_{l,m}(\kappa )=\kappa \, \}.\]

\begin{lem}
The map
\[
   (E^{\otimes (l+m)})^{*}_{\mathrm{sym}(l,m)}\ni \kappa \mapsto
    \Xi _{l,m}(\kappa )\in \mathcal{L}((E),(E)^{*})
\]
is injective. Moreover, for \( \kappa \in (E^{\otimes (l+m)})^{*}_{\mathrm{sym}(l,m)} \)
and \( \kappa \in (E^{\otimes (l'+m')})^{*}_{\mathrm{sym}(l',m')} \), if \( \Xi _{l,m}(\kappa )=\Xi _{l',m'}(\kappa ') \),
then \( l=l' \), \( m=m' \) and \( s_{l,m}(\kappa )=s_{l,m}(\kappa ') \).
\end{lem}

Due to this lemma, the map
\begin{equation}
\bigoplus _{l,m=0}^{\infty }(E^{\otimes (l+m)})^{*}_{\mathrm{sym}(l,m)}
    \ni \{\kappa _{l,m}\}_{l,m=0}^{\infty }\mapsto
    \sum _{l,m=0}^{\infty }\Xi _{l,m}(\kappa _{l,m})\in \mathcal{L}((E),(E)^{*})
\label{eq:symm_(l,m)<-1to1->IntegralKernelOp}
\end{equation}
is injective. 

\begin{prop}
[\bf Fock expansion]\label{thm:Fock_expansion}
The map \eqref{eq:symm_(l,m)<-1to1->IntegralKernelOp}
is surjective, i.e., for any \( \Xi \in \mathcal{L}((E),(E)^{*}) \), there
exists an unique \( \{\kappa _{l,m}\}_{l,m=0}^{\infty } \), \( \kappa _{l,m}\in (E^{\otimes (l+m)})^{*}_{\mathrm{sym}(l,m)} \)
such that 
\begin{equation}
\Xi \phi =\sum _{l,m=0}^{\infty }\Xi _{l,m}(\kappa _{l,m})\phi ,\quad \phi \in (E)
\label{eq:Fock_expansion}
\end{equation}
where the sum of the right hand side of \eqref{eq:Fock_expansion} converges in
\( (E)^{*} \). 

If \( \Xi \in \mathcal{L}((E),(E)) \), then
\[
\kappa _{l,m}\in E^{\widehat{\otimes }l}\otimes \left( E^{\otimes m}\right) ^{*}_{\mathrm{sym}},\quad l,m\geq 0\]
and the sum of the right hand side of \eqref{eq:Fock_expansion} converges in
\( (E) \).
\end{prop}
\begin{proof}
See section 4.4 and 4.5 of \cite{Obata}.
\end{proof}

\begin{prop}
\label{thm:adjoint_of_integral_kernel_op}
For \( \Xi _{l,m}(\kappa _{l,m})\in \mathcal{L}((E),(E)^{*}) \)
it follows that
\[
\Xi _{l,m}(\kappa _{l,m})^{*}=\Xi _{m,l}(t_{m,l}(\kappa _{l,m}))\]
where the map \( t_{m,l} \) is defined by
\[
\left\langle t_{m,l}(\kappa _{l,m}),\eta \otimes \zeta \right\rangle 
   :=\left\langle \kappa _{l,m},\zeta \otimes \eta \right\rangle ,\quad
       \eta \in E^{\otimes m},\, \zeta \in E^{\otimes l}.\]

\end{prop}

By the way, for \( \Xi _{l,m}(\kappa )\in \mathcal{L}((E),(E)^{*}) \) and \( \Xi _{l',m'}(\lambda )\in \mathcal{L}((E),(E)) \),
we have \( \Xi _{l,m}(\kappa )\Xi _{l',m'}(\lambda )\in \mathcal{L}((E),(E)^{*}) \).
From proposition \ref{thm:Fock_expansion}, we can infer that \( \Xi _{l,m}(\kappa )\Xi _{l',m'}(\lambda ) \)
is expressed as a sum of integral kernel operators.

\begin{defn}
\label{thm:S^l_(m-k)^l'_(m'-k)}Let \( \kappa \in \left( E^{\otimes (l+m)}\right) ^{*} \),
\( \lambda \in E^{\otimes l'}\otimes \left( E^{\otimes m'}\right) ^{*} \) and
\( m\wedge l':=\min \{m,l'\} \). For \( 0\leq k\leq m\wedge l' \), we define
\( S_{m-k}^{l}\, _{m'}^{l'-k}(\kappa \circ _{k}\lambda )\in
\left( E^{\otimes (l+l'+m+m'-2k)}\right) ^{*} \)
as follows.
\[
\begin{split}
S_{m-k}^{l} \, _{m'}^{l'-k}(\kappa \circ _{k}\lambda )
  :=&\sum _{\mathbf{i},\mathbf{j},\mathbf{i}',\mathbf{j}'}
     \sum _{\mathbf{h}}\left\langle \kappa ,
     e(\mathbf{i})\otimes e(\mathbf{j})\otimes e(\mathbf{h})\right\rangle \\
  & \times \left\langle \lambda ,e(\mathbf{h})\otimes 
    e(\mathbf{i}')\otimes e(\mathbf{j}')\right\rangle
  e(\mathbf{i})\otimes e(\mathbf{i}')\otimes
     e(\mathbf{j})\otimes e(\mathbf{j}'),
\end{split}
\]
where \( \mathbf{i} \), \( \mathbf{j} \), \( \mathbf{i}' \), \( \mathbf{j}' \),
and \( \mathbf{h} \) runs over the whole of \( \mathbf{N}^{l} \), \( \mathbf{N}^{m-k} \),
\( \mathbf{N}^{l'-k} \), \( \mathbf{N}^{m'} \), and \( \mathbf{N}^{k} \)
respectively.
\end{defn}

We check well-definedness of \( S_{m-k}^{l}\, _{m'}^{l'-k}(\kappa \circ _{k}\lambda ) \).
Now there exists \( p\geq 0 \) such that \( |\kappa |_{-p}=|\kappa |_{l,m;-p,-p}<+\infty  \).
Note that for any \( p'\geq 0 \) there exists \( q'=q'(l',m';p')\geq 0 \)
such that \( |\lambda |_{l',m';p',-(p'+q')}<\infty  \). Let \( p':=p \) and
\( q:=q'(l',m';p) \). Then
\[
\begin{split}
\big| S&_{m-k}^{l}\, _{m'}^{l'-k}(\kappa \circ _{k}\lambda )
   \big| ^{2}_{l+l'-k,m+m'-k;-p,-(p+q)} \\
 & \leq \sum _{\mathbf{i},\mathbf{j},\mathbf{i}',\mathbf{j}'}
   \sum _{\mathbf{h}}\left| \left\langle \kappa ,e(\mathbf{i})
     \otimes e(\mathbf{j})\otimes e(\mathbf{h})
     \right\rangle \right| ^{2}\left| e(\mathbf{h})\right| ^{2}_{-p} \\
 & \quad \times \sum _{\mathbf{h}'}\left| \left\langle \kappa ,e(\mathbf{h}')
     \otimes e(\mathbf{i}')\otimes e(\mathbf{j}')
     \right\rangle \right| ^{2}\left| e(\mathbf{h}')\right| ^{2}_{p}
     \left| e(\mathbf{i})\otimes e(\mathbf{i}')\right| _{-p}
     \left| e(\mathbf{j})\otimes e(\mathbf{j}')\right| _{-(p+q)}.
\end{split}
\]
Since \( \inf \mathrm{Spec}(A)>1 \), we have
\[
\left| e(\mathbf{j})\right| _{-(p+q)}=\left| e(\mathbf{j})\right| _{-p}\left| e(\mathbf{j})\right| _{-q}\leq \left| e(\mathbf{j})\right| _{-p}\]
and
\( \left| e(\mathbf{i}')\right| _{-p}\leq \left| e(\mathbf{i}')\right| _{p} \). 
This implies that
\[
\left| S_{m-k}^{l}\, _{m'}^{l'-k}(\kappa \circ _{k}\lambda )\right| _{l+l'-k,m+m'-k;-p,-(p+q)}\leq |\kappa |_{-p}|\lambda |_{l',m';p,-(p+q)}<+\infty .\]

\begin{prop}
For \( \Xi _{l,m}(\kappa )\in \mathcal{L}((E),(E)^{*}) \)
and \( \Xi _{l',m'}(\lambda )\in \mathcal{L}((E),(E)) \),
it follows that
\[
\Xi _{l,m}(\kappa )\Xi _{l',m'}(\lambda )
=\sum _{k=0}^{m\wedge l'}k!\left( \begin{array}{c}
m\\
k
\end{array}\right) \left( \begin{array}{c}
l'\\
k
\end{array}\right)
\Xi _{l+l'-k,m+m'-k}(S_{m-k}^{l}\, _{m'}^{l'-k}(\kappa \circ _{k}\lambda )).\]

\end{prop}
The formal part of the proof of this proposition is the same as proposition
7.3 of \cite{Ji-Obata}. From definition \ref{thm:S^l_(m-k)^l'_(m'-k)}, the
analytic part of the proof of this proposition is obvious.

Finally, we define an useful tool for an analysis of a Fock space.

\begin{defn}
For \( f\in H \), 
\[
\exp (f):=\sum _{n=0}^{\infty }\frac{1}{n!}f^{\otimes n}\in \Gamma _{\mathrm{b}}(H)\]
is called an {\it exponential vector} or a {\it coherent vector}.
\end{defn}
The following lemmas are well-known facts.

\begin{lem}
\( \{\exp (f)\, |\, f\in H\} \) generates \( \Gamma _{\mathrm{b}}(H) \).
Moreover, \( \{\exp (f)\, |\, f\in E\} \) spans a dense subspace of \( (E) \).
\end{lem}

\section{The gauge group and its representation}\label{section:construction-energy-rep}

Let \( G \) be a Lie group and \( \mathfrak {g} \) be the Lie algebra of \( G \)
and \( \mathfrak {g}^{\mathrm{c}} \) be the complexification of \( \mathfrak {g} \).
Let ``\( \mathrm{Ad} \)'' be the adjoint representation of \( G \). For
the complexification of ``\( \mathrm{Ad} \)'', we use the same notation.

Now let \( G \) be a semi-simple compact Lie group. Then the Killing form \( K \)
on \( \mathfrak {g} \) is negative-definite. Thus \( (-K) \) is an inner product
on \( \mathfrak {g} \) and hence we can define an inner product \( (\cdot ,\cdot )_{\mathfrak {g}} \)
of \( \mathfrak {g}^{\mathrm{c}} \) via the polarization identity.
The representation \( (\mathfrak {g}^{\mathrm{c}},\mathrm{Ad}) \) of \( G \)
is a unitary representation with respect to the inner product \( (\cdot ,\cdot )_{\mathfrak {g}} \). 

For \( X, \) \( Y\in \mathfrak {g} \), let 
\[
\mathrm{ad}(X)Y:=[X,Y]\]
where \( [X,Y] \) be the Lie bracket of \( \mathfrak {g} \). Then \( \mathrm{ad}(X) \)
is a representation of the Lie algebra \( \mathfrak {g} \) on the vector space
\( \mathfrak {g} \). The following lemma is a well-known fact about the Lie
algebra \( \mathfrak {g} \) and its Killing form. 

\begin{lem}
\label{thm:adjoint_op_of_ad(Z)}Let \( \mathfrak {g} \) be a Lie algebra, and
\( K \) be the Killing form on \( \mathfrak {g} \). Then
\[
K(X,\mathrm{ad}(Z)Y)=-K(\mathrm{ad}(Z)X,Y)\]
for all \( X,Y,Z\in \mathfrak {g} \). Moreover, for \( X \), \( Y \), 
\( Z\in \mathfrak {g}^{\mathrm{c}} \),
we have
\[
(\mathrm{X},\mathrm{ad}(Z)Y)_{\mathfrak {g}}
  =(-\mathrm{ad}(\overline{Z})X,Y)_{\mathfrak {g}}\]
where
\[
\overline{Z_{1}+\sqrt{-1}Z_{2}}:=Z_{1}-\sqrt{-1}Z_{2}.\]
for \( Z_{1} \), \( Z_{2}\in \mathfrak {g} \). 
\end{lem}
We shall use this lemma in the following section.

Next, we define the gauge group and its representation. Let \( M \) be a
Riemannian manifold without boundary
and \( (\cdot ,\cdot )_{x} \) be a inner product on \( T_{x}^{*}M \)
determined by the Riemannian structure of \( M \). 

Let \( C^{\infty }_{\mathrm{c}}(M,G) \) be the set of all \( C^{\infty } \)-mappings
\( \psi :M \rightarrow G \) with compact support. 
We call \( C^{\infty }_{\mathrm{c}}(M,G) \)
the gauge group. Let \( C^{\infty }_{\mathrm{c}}(M,\mathfrak {g}) \) be the set
of all \( C^{\infty } \)-mappings \( \Psi : M\rightarrow \mathfrak {g} \)
with compact support.
This is the ``Lie algebra'' of \( C^{\infty }_{\mathrm{c}}(M,G) \).

Let \( \Omega ^{1}(M) \) be the space of real-valued 
1-forms on \( M \) with compact support and \( \Omega ^{1}(M,\mathfrak {g})
:=\Omega ^{1}(M)\otimes \mathfrak {g}^{\mathrm{c}} \).
We can define a natural inner product on \( \Omega ^{1}(M,\mathfrak {g}) \)
as follows. First, let 
\[
  \left\langle \omega _{x}\otimes X,\omega '_{x}\otimes X'\right\rangle _{x}
   :=(\omega _{x},\omega '_{x})_{x}(X,X')_{\mathfrak {g}}
\]
for all \( \omega _{x} \), \( \omega '_{x}\in T_{x}^{*}M \) and \( X \),
\( X'\in \mathfrak {g}^{\mathrm{c}} \). For each \( x\in M \), \( \left\langle \cdot ,\cdot \right\rangle _{x} \)
is an inner product on \( T^{*}_{x}M\otimes \mathfrak {g}^{\mathrm{c}} \).
Then
\begin{equation}
\left\langle f,g\right\rangle _{0}:=\int _{M}\left\langle f(x),g(x)\right\rangle _{x}dv,
\label{eq:inner_prod_of_H(M,g)}
\end{equation}
where \( dv \) is the volume measure on \( M \), and where \( f \), \( g\in \Omega ^{1}(M,\mathfrak {g}) \).
This is an inner product on \( \Omega ^{1}(M,\mathfrak {g}) \). We denote the
completion of \( \Omega ^{1}(M,\mathfrak {g}) \) with respect to the inner
product \( \left\langle \cdot ,\cdot \right\rangle _{0} \) by \( H(M,\mathfrak {g}) \). 

Let
\[
   (V(\psi )f)(x):=[\mathrm{id}_{T_{x}^{*}M}\otimes \mathrm{Ad}(\psi (x))]f(x), \quad x\in M
\]
for all \( \psi \in C^{\infty }_{\mathrm{c}}(M,G) \) and \( f\in H(M,\mathfrak {g}) \).
Then \( V(\psi ) \) is a unitary operator on the Hilbert space \( H(M,\mathfrak {g}) \).
We call \( V \) the {\it adjoint representation} of
gauge group \( C_{\mathrm{c}}^{\infty }(M,G) \).

For \( \psi \in C^{\infty }_{\mathrm{c}}(M,G) \), we define the right logarithmic
derivative \( \beta (\psi )\in \Omega ^{1}(M,\mathfrak {g}) \) as follows.
\[
(\beta (\psi ))(x):=(d\psi )_{x}\psi (x)^{-1}.\]
\( \beta (\psi ) \) satisfies
\begin{equation}
\label{eq:Maurer-Cartan_cocycle}
\beta (\psi \cdot \varphi )=V(\psi )\beta (\varphi )+\beta (\psi ),
\end{equation}
where \( \psi \cdot \varphi  \) is defined by the pointwise multiplication.
The relation \eqref{eq:Maurer-Cartan_cocycle} is called the {\it Maurer-Cartan cocycle}.

\begin{defn}
Let \( U(\psi ) \) be an unitary representation on the on the Boson Fock space \( \Gamma _{\mathrm{b}}(H(M,\mathfrak {g})) \)
determined by 
\[
     U(\psi )\exp (f):=
       \exp \left( -\frac{1}{2}|\beta (\psi )|^{2}_{0} \right)
       \exp \left( -\left\langle 
            \beta (\psi ),V(\psi )f
       \right \rangle _{0}\right) 
       \exp \left( V(\psi )f+\beta (\psi )\right)
\]
for \( f\in H(M,\mathfrak {g}) \) and \( \psi \in C^{\infty }_{\mathrm{c}}(M,G) \).
We call \( U \) the {\it energy representation} of the gauge group \( C^{\infty }_{\mathrm{c}}(M,G) \).
\end{defn}

Now we construct a CH-space \( E \) by using the Hilbert space \( H(M,\mathfrak {g}) \) 
and a self-adjoint operator on \( H(M,\mathfrak {g})\). 
Let \( M \) be a compact Riemann manifold without boundary. 
Let \( \Lap  \) be the Bochner Laplacian on \( \Omega ^{1}(M) \) 
determined by the Levi-Civita connection on \( M \) 
and \( H(M) \) be the completion of \( \Omega ^{1}(M) \). Then \( \Lap +2 \) is an essentially
self-adjoint operator on \( \Omega ^{1}(M) \). (This is shown
by considering the complexification of \( \Lap +2 \).) 
Let \( A \) be a closed extension of
\( (\Lap +2)\otimes \mathrm{id}_{\mathfrak {g}} \). Then there
exists a C.O.N.S. \( \{e_{i}\}_{i\in \mathbf{N}} \) of \( H(M) \) consisting
of eigenvectors of the essentially self-adjoint operator \( \Lap +2 \) on \( \Omega ^{1}(M) \).
For any C.O.N.S. \( \{u_{j}\}_{j=1}^{\dim \mathfrak {g}} \) of \( \mathfrak {g^{\mathrm{c}}} \),
\begin{equation}
  \{ e(i,j):=e_{i}\otimes u_{j}\, |\, 
   i\in \mathbf{N},\, 1\leq j\leq \dim \mathfrak {g}\}
  \label{eq:CONS_of_H(M,g)}
\end{equation}
is a C.O.N.S. of \( H(M,\mathfrak {g}) \).
This C.O.N.S. of \( H(M,\mathfrak {g}) \) and the essentially self-adjoint operator \( A \)
satisfy the condition (i), and (ii) of definition \ref{thm:property_of_self-adj_op_A}
and \eqref{eq:Hilbert-Schmidt_inverse}.
(As for the general theory of Laplacian on a vector bundle, see chapter 1 of
\cite{Gilkey}.) For the sake of the calculation in section \ref{section:proof-of-irreducibility},
we take a C.O.N.S. of \( \mathfrak {g}^{\mathrm{c}} \) as follows.

Let \( \Delta  \) be a root system of \( \mathfrak {g}^{\mathrm{c}} \) and
\( \Delta ' \) be a set of all positive roots of \( \Delta  \).
Let \( \mathfrak {h} \) be a Cartan subalgebra of \( \mathfrak {g} \) and
\( \{H_{1},\ldots ,H_{\dim \mathfrak {h}}\} \) be a C.O.N.S. of \( \mathfrak {h} \).
Let \( X_{\alpha } \), \( \alpha \in \Delta  \) be a normalized element of
\( \mathfrak {g}^{\mathrm{c}} \) such that \( [H,X_{\alpha }]=\alpha (H)X_{\alpha } \)
for all \( H\in \mathfrak {h} \). Then we have a C.O.N.S.
\begin{equation}
   \{H_{1},\ldots ,H_{\dim \mathfrak {h}},X_{\alpha },X_{-\alpha }\, |\, \alpha \in \Delta '\}
\label{eq:CONS_of_g}
\end{equation}
of a complex vector space \( \mathfrak {g}^{\mathrm{c}} \) with respect to
the inner product on \( \mathfrak {g}^{\mathrm{c}} \). 

Thus we obtain the CH-space \( E \) constructed from \( (H(M,\mathfrak {g}), A) \).

Next, we construct a CH-space \( (E) \) by using
the Hilbert space \( \Gamma _{\mathrm{b}} (H(M,\mathfrak {g}))\)
and a self-adjoint operator on \( \Gamma _{\mathrm{b}} (H(M,\mathfrak {g}))\). 
We write a C.O.N.S. of \( H(M,\mathfrak {g})^{\widehat{\otimes }n} \)
\( (n\geq 1) \) in terms of \( H(M,\mathfrak {g}) \). 
Put \( \dim \mathfrak {g}=N_{0} \), \( \dim \mathfrak {h}=N_{1} \), and
\[
\Delta '=\{\alpha _{1},\ldots ,\alpha _{N_{2}}\},\quad (N_{2}:=\#\Delta '=\frac{1}{2}(N_{0}-N_{1})).\]
Let
\[
u_{j}:=\left\{ \begin{array}{ll}
H_{j}, & \mathrm{if}\, \, 1\leq j\leq N_{1},\\
X_{\alpha _{j-N_{1}}}, & \mathrm{if}\, \, N_{1}+1\leq j\leq N_{1}+N_{2},\\
X_{-\alpha _{j-(N_{1}+N_{2})}}, & \mathrm{if}\, \, N_{1}+N_{2}+1\leq j\leq N_{1}+2N_{2}=N_{0}
\end{array}\right. \]

Fix \( d\in \{1,2,\ldots ,n\} \). Let 
\begin{gather*}
\mathbf{i}:=(\overbrace{i_{1},\ldots ,i_{1}}^{N(1)\, times},
            \overbrace{i_{2},\ldots ,i_{2}}^{N(2)\, times},\ldots ,
            \overbrace{i_{d},\ldots ,i_{d}}^{N(d)\, times})\in \mathbf{N}^{n}, \\
N(1)+N(2)+\ldots +N(d)=n,\quad i_{1}<i_{2}<\ldots <i_{d}. 
\end{gather*}
For this \( \mathbf{i}\in \mathbf{N}^{n} \), we define 
\( \mathbf{j}\in \{1,2,\ldots ,N_{0}=\dim \mathfrak {g} \} \)
as follows.
\[
\mathbf{j}=(j(i_{1},1),\ldots ,j(i_{1},N(1)),j(i_{2},1),\ldots ,j(i_{2},N(2)),j(i_{d},1),\ldots ,j(i_{d},N(d)))\]
where \( j(i,k)\in \{1,2,\ldots ,N_{0}\} \) satisfies the following conditions:
for each \( i\in \mathbf{N} \)
if \( k_{1}<k_{2} \), then \( j(i,k_{1})\le j(i,k_{2}) \). 

Let \( \Lambda (n) \) be the subset of \( \mathbf{N}^{n}\times \{1,2,\ldots ,N_{0}\}^{n} \)
which consists of all \( (\mathbf{i},\mathbf{j}) \). For \( (\mathbf{i},\mathbf{j})\in \Lambda (n) \), let
\[
\begin{split}
\widehat{e}(\mathbf{i},\mathbf{j}):=
&e(i_{1},j(i_{1},1))\widehat{\otimes }\ldots
   \widehat{\otimes }e(i_{1},j(i_{1},N(1))) \\
& \widehat{\otimes }e(i_{2},j(i_{2},1))\widehat{\otimes }\ldots
   \widehat{\otimes }e(i_{2},j(i_{2},N(2))) \\
& \ldots \widehat{\otimes }e(i_{d},j(i_{d},1))\widehat{\otimes }\ldots
   \widehat{\otimes }e(i_{d},j(i_{d},N(d))),
\end{split}
\]
then \( \{\widehat{e}(\mathbf{i},\mathbf{j})\, |\, (\mathbf{i},\mathbf{j})\in \Lambda (n)\} \)
is a C.O.N.S. of \( H(M,\mathfrak {g})^{\widehat{\otimes }n} \), \( n\geq 1\). 
This C.O.N.S. of \( \) of \( \Gamma _{\mathrm{b}}(H(M,\mathfrak {g}))\)
and the essentially self-adjoint operator \( \Gamma _{\mathrm{b}}(A) \)
satisfy the condition (i), and (ii) of definition \ref{thm:property_of_self-adj_op_A}
and \eqref{eq:Hilbert-Schmidt_inverse}.

Therefore we obtain a CH-space \( (E) \) constructed from
\( (\Gamma _{\mathrm{b}}(H(M,\mathfrak {g})),\Gamma _{\mathrm{b}}(A)) \).

In this paper, we discuss irreducibility of the energy representation \( U \)
of \( C^{\infty }(M,G) \) with the help of the white noise calculus,
however, it is difficult to deal with the energy representation \( U \) directly.
Thus we treat not the representation of ``Lie group'' \( C^{\infty }(M,G) \)
but the representation of ``Lie algebra'' \( C^{\infty }(M,\mathfrak {g}) \). 

We introduce a proposition for the differentiability of a operator \( V(\psi ) \)
on the CH-space \( E \). 

\begin{prop}
\label{thm:1-para-subgroup_differentiability}
Let \( \psi _{t}(x):=\exp \left( t\, \Psi (x)\right)  \)
for \( \Psi \in C^{\infty }(M,\mathfrak {g}) \) and \( t\in \mathbf{R} \).
Then \( \left\{ V(\psi _{t})\right\} _{t\in \mathbf{R}} \) is a regular one-parameter
subgroup of \( GL(E) \), namely, for any \( p\geq 0 \) there exists \( q\geq 0 \)
such that
\[
  \lim _{t\rightarrow 0}\sup _{f\in E;\left| f\right| _{q}\leq 1}\left|
   \frac{V(\psi _{t})f-f}{t}-V(\Psi )f\right| _{p}=0,
\]
where
\[
  (V(\Psi )f)(x):=[\mathrm{id}_{T_{x}^{*}M}\otimes \mathrm{ad}(\Psi (x))]f(x),
  \quad x\in M
\]
for all \( f\in E \). 
\end{prop}

\begin{proof}
Note that there exists \( C(\Psi ,p)>0 \) such that
\[
   \left\Vert V(\Psi )f\right\Vert _{p}
     \leq C(\Psi ,p)\left\Vert f\right\Vert _{p},
         \quad f\in \Omega ^{1}(M,\mathfrak {g})
\]
for each \( \Psi \in C^{\infty }(M,\mathfrak {g}) \)
and \( p\in \mathbf{N} \), i.e. \( V(\Psi )\in \mathcal{L}(E,E) \).
(See proposition 2.5 of \cite{Samelson}.)
Since
\[
\begin{split}
(V(\psi _{t})f)(x) 
&= [\mathrm{id}_{T_{x}^{*}M} \otimes \mathrm{Ad}(\exp (t\, \Psi (x))) ]f(x)\\
&= [\mathrm{id}_{T_{x}^{*}M} \otimes \exp (t\, \mathrm{ad}(\Psi (x)))]f(x)\\
&= \left[ \mathrm{id}_{T_{x}^{*}M} \otimes 
   \sum _{k=0}^{\infty }\frac{1}{k!}(t\, \mathrm{ad}(\Psi (x)))^{k}
   \right] f(x) \\
&= \left( \sum _{k=0}^{\infty }\frac{1}{k!}(t\, V(\Psi ))^{k}f\right) (x)
\end{split}
\]
for each \( x\in M \), we have
\[
\begin{split}
\Bigg\Vert \frac{V(\psi _{t})f-f}{t}-V(\Psi )f \Bigg\Vert _{p}
& \leq \frac{1}{t}\sum _{k=2}^{\infty }\frac{1}{k!}
    \left\Vert (t\, V(\Psi ))^{k}f\right\Vert _{p}
  \leq \frac{1}{t}\sum _{k=2}^{\infty }\frac{1}{k!}
    (t\, C(\Psi ,p))^{k}\left\Vert f\right\Vert _{p}\\
& \leq t\left( \sum _{k=0}^{\infty }\frac{1}{k!}C(\Psi ,p)^{k}\right)
  \left\Vert f\right\Vert _{p}
  = t\exp (C(\Psi ,p))\left\Vert f\right\Vert _{p}
\end{split}
\]
for \( 0<t<1 \). This implies that
\[
  \lim _{t\rightarrow 0} \sup _{\left\Vert f\right\Vert _{p}
   \leq 1}\left\Vert \frac{V(\psi _{t})f-f}{t}-V(\Psi )f\right\Vert _{p}=0.
\]
\end{proof} 

This proposition
plays a crucial role of our proof of the differentiability of the energy representation
\( U(\psi _{t}) \), that is, 

\begin{lem}
Let \( \psi _{t}(x):=\exp \left( t\Psi (x)\right)  \) 
for \( \Psi \in C^{\infty }(M,\mathfrak {g}) \)
and \( t\in \mathbf{R} \). Then \( \{U(\psi _{t})\}_{t\in \mathbf {R}} \) is
a regular one-parameter subgroup of \( GL((E)) \) with infinitesimal generator
\( d\Gamma _{\mathrm{b}}(V(\Psi ))+D_{d\Psi }^{*}-D_{d\Psi } \).
\end{lem}
\begin{proof}
Let \( \mu  \) be the Gaussian measure on the real vector space \( E^{*}_{\mathrm {Re}} \)
of distributions. Then the Boson Fock space is identified with the space of complex valued
\( L^{2} \)-functions on \( E^{*}_{\mathrm {Re}} \) with respect to \( \mu  \).
For \( \eta \in E_{\mathrm {Re}}^{*} \), let 
\[
 (T_{\eta }\phi )(x)
   :=\exp \left( -\frac{1}{2}|\eta |^{2}_{0}\right)
     \phi _{\eta }(x)\phi (x-2\eta ),\quad
       \phi \in L^{2}( E^{*}_{\mathrm {Re}},\mathbf{C} ; \mu).
\]
Here
\[
  \phi _{\eta }(x)
    :=\sum _{n=0}^{\infty }\frac{1}{n!}
       \left\langle :x^{\otimes n}:,\eta ^{\otimes n}\right\rangle ,\quad
       x\in E^{*}_{\mathrm{Re}}.
\]
Then \( T_{\beta (\psi) } \) satisfies
\( U(\psi )=T_{\beta (\psi )}\Gamma _{\mathrm {b}}(V(\psi )) \) and
\( T_{\beta (\psi _{t})}=T_{td\Psi } \) is
a regular one-parameter subgroup of \( GL((E)) \) with infinitesimal generator
\( D_{d\Psi }^{*}-D_{d\Psi } \). (See section 5.7 of \cite{Obata}.)
Moreover, from proposition 5.4.5 of \cite{Obata} and proposition
\ref{thm:1-para-subgroup_differentiability},
\( \Gamma _{\mathrm {b}}(V(\psi _{t})) \) is a regular one-parameter subgroup
of \( GL((E)) \) with infinitesimal generator \( d\Gamma _{\mathrm {b}}(V(\Psi )) \).
Thus \( U(\psi _{t}) \) is a regular one-parameter subgroup of \( GL((E)) \)
with infinitesimal generator \( d\Gamma _{\mathrm{b}}(V(\Psi ))+D_{d\Psi }^{*}-D_{d\Psi } \).
\end{proof}

Let \( \pi (\Psi ):=d\Gamma _{\mathrm{b}}(V(\Psi ))+D_{d\Psi }^{*}-D_{d\Psi } \).
Then \( \pi (\Psi )\in \mathcal{L}((E),(E)) \) and \( \pi (\Psi ) \) has the
following expression :
\[
\pi (\Psi )=\Xi _{1,1}((\mathrm{id}\otimes V(\Psi ))^{*}\tau )+\Xi _{1,0}(d\Psi )-\Xi _{0,1}(d\Psi ),\]
where \( \tau \in (E\otimes E)^{*} \) is defined by
\[
\left\langle \tau ,f\otimes g\right\rangle :=\left\langle f,g\right\rangle ,\quad f,g\in E.\]
(See proposition 4.5.3 of \cite{Obata}.) To avoid notational complexity, 
we put \( \lambda _{1,1}=(\mathrm{id}\otimes V(\Psi ))^{*}\tau  \)
and \( \lambda _{1,0}=\lambda _{0,1}=d\Psi  \)
for \( \Psi \in C^{\infty }(M,\mathfrak {g}) \), namely
\[
  \pi (\Psi )=\Xi _{1,1}(\lambda _{1,1})+\Xi _{1,0}(\lambda _{1,0})
               -\Xi _{0,1}(\lambda _{0,1}).
\]
Let 
\[
  \widetilde{\pi }(\Psi ):=-\pi (\Psi )^{*},
   \quad \Psi \in C^{\infty }(M,\mathfrak {g}).
\]
Then \( \widetilde{\pi }(\Psi )\in \mathcal{L}((E)^{*},(E)^{*}) \). Moreover,
we can check that \( \pi (\Psi )^{*}|(E)=-\pi (\Psi ) \) by using proposition
\ref{thm:adjoint_of_integral_kernel_op} and lemma \ref{thm:adjoint_op_of_ad(Z)}.
In fact, since we have \( t_{1,1}(\lambda _{1,1})=-\lambda _{1,1} \), it satisfies
that \( \Xi _{1,1}(\lambda _{1,1})^{*}|(E)=-\Xi _{1,1}(\lambda _{1,1}) \).

\section{Irreducibility of the energy representation}\label{section:proof-of-irreducibility}

We now give the main theorem of this paper. 

\begin{thm}\label{thm:main-theorem}
Let \( M \) be a compact Riemannian manifold without boundary. 
Then the energy representation \( \{U(\psi )|\psi \in C^{\infty }(M,G)\} \) is irreducible.
\end{thm}
Let \( \Xi \in \mathcal{L}(\Gamma _{\mathrm{b}}(H(M,\mathfrak {g})), 
\Gamma _{\mathrm{b}}(H(M,\mathfrak {g}))) \)
satisfy
\begin{equation}
\label{eq:U(psi)_commute_Theta}
U(\exp (t\Psi ))\Xi =\Xi U(\exp (t\Psi ))
\end{equation}
for all \( \Psi \in C^{\infty }(M,\mathfrak {g}) \), \( t\in \mathbf{R} \).
Note that the restriction \( \Xi |(E) \) of \( \Xi \) to \( (E) \) is
a continuous linear operator from \( (E) \) to \( (E)^{*} \) and \( \pi (\Psi )^{*} \)
is a continuous linear operator on \( (E)^{*} \). Then \eqref{eq:U(psi)_commute_Theta}
implies
\begin{equation}
\widetilde{\pi }(\Psi )\Xi =\Xi \pi (\Psi )
\label{eq:pi(Psi)_commute_Theta}
\end{equation}
for all \( \Psi \in C^{\infty }(M,\mathfrak {g}) \) as a continuous linear
operator from \( (E) \) to \( (E)^{*} \). Our main problem is to find \( \Xi  \)
satisfying \eqref{eq:pi(Psi)_commute_Theta}. 

\begin{lem}
Let
\[
 \Xi =\sum _{l,m=0}^{\infty } \Xi _{l,m}(\kappa _{l,m}).
\]
be the Fock expansion for a continuous linear operator \( \Xi \)
from \( (E) \) to \( (E)^{*} \). 
If \( \Xi \) satisfies \eqref{eq:pi(Psi)_commute_Theta}, then
\( \kappa _{l,m}\), \( l,m \geq 0\) satisfy the following relations:
\begin{equation}
\begin{split}
s_{l,0}&\left( S_{0}^{l}\, _{0}^{0}(\kappa _{l,1}\circ _{1}\lambda _{1,0})\right) \\
  &=s_{l,0}\left( l\, S_{0}^{1}\, _{0}^{l-1}(\lambda _{1,1}\circ _{1}\kappa _{l,0})
   -(l+1)S_{0}^{0}\, _{0}^{l}(\lambda _{0,1}\circ _{1}\kappa _{l+1,0})\right) ,
\label{eq:irr-iff-condition(2)}
\end{split}
\end{equation}
\begin{equation}
\begin{split}
s_{0,m}&\Big( -S_{0}^{0}\, _{m}^{0}(\lambda _{0,1}\circ _{1}\kappa _{1,m})\Big) \\
 &=s_{0,m}\Big( m\, S_{m-1}^{0}\, ^{0}_{1}(\kappa _{0,m}\circ _{1}\lambda _{1,1})
  +(m+1)\, S_{m}^{0}\, _{0}^{0}(\kappa _{0,m+1}\circ _{1}\lambda _{1,0})\Big) ,
\label{eq:irr-iff-condition(3)}
\end{split}
\end{equation}
\begin{equation}
\begin{split}
s_{l,m}&\left( m\, S_{m-1}^{l}\, _{1}^{0}(\kappa _{l,m}\circ _{1}\lambda _{1,1})
  +(m+1)S_{m}^{l}\, _{0}^{0}(\kappa _{l,m+1}\circ _{1}\lambda _{1,0})\right)\\
&=s_{l,m}\left( l\, S_{0}^{1}\, _{m}^{l-1}(\lambda _{1,1}\circ _{1}\kappa _{l,m})
  -(l+1)S_{0}^{0}\, _{m}^{l}(\lambda _{0,1}\circ _{1}
   \kappa _{l+1,m})\right) \label{eq:irr-iff-condition(4)}
\end{split}
\end{equation}
for all \( l,m\geq 1 \) and
\begin{equation}
-\lambda _{0,1}\circ _{1}\kappa _{1,0}=\kappa _{0,1}\circ _{1}\lambda _{1,0}.
\label{eq:irr-iff-condition(1)}
\end{equation}
\end{lem}

\begin{proof} 
\[
\begin{split}
&\left( \sum _{l,m=0}^{\infty }\Xi _{l,m}
  (\kappa _{l,m})\right) \Xi _{1,0}(\lambda _{1,0}) \\
& =\sum _{l,m=0}^{\infty }\sum _{k=0}^{m\wedge 1}k!
  \dbinom{m}{k} \dbinom{1}{k}
  \Xi _{l+1-k,m-k}\left( S_{m-k}^{l}\, _{0}^{1-k}
  (\kappa _{l,m}\circ _{k}\lambda _{1,0}\right)  \\
& =\Xi _{0,0}(\kappa _{0,1}\circ _{1}\lambda _{1,0}) \\
& \quad +\sum _{l=1}^{\infty }\Xi _{l,0}\left( S_{0}^{l-1}\, _{0}^{1}
  (\kappa _{l-1,0}\circ \lambda _{1,0}+S_{0}^{l}\, _{0}^{0}
  (\kappa _{l,1}\circ _{1}\lambda _{1,0})\right)  \\
& \quad +\sum ^{\infty }_{m=1}\Xi _{0,m}\left( (m+1)S_{m}^{0}\, _{0}^{0}
  (\kappa _{0,m+1}\circ _{1}\lambda _{1,0})\right)  \\
& \quad +\sum _{l,m=1}^{\infty }\Xi _{l,m}\left( S_{m}^{l-1}\, _{0}^{1}
  (\kappa _{l-1,m}\circ \lambda _{1,0})+(m+1)S_{m}^{l}\, _{0}^{0}
  (\kappa _{l,m+1}\circ _{1}\lambda _{1,0})\right)
\end{split}
\]
and
\[
\begin{split}
&\left( \sum _{l,m=0}^{\infty }\Xi _{l,m}(\kappa _{l,m})\right) \Xi _{0,1}(\lambda _{0,1}) \\
&= \sum _{l,m=0}^{\infty }\Xi _{l,m+1}\left( S_{l}^{m}\, _{1}^{0}
   (\kappa _{l,m}\circ \lambda _{0,1})\right) \\
&= \sum _{m=1}^{\infty }\Xi _{0,m}\left( S_{m-1}^{0}\, _{1}^{0}
   (\kappa _{0,m-1}\circ \lambda _{0,1})\right) +\sum _{l,m=1}^{\infty }
   \Xi _{l,m}\left( S_{m-1}^{l}\, _{1}^{0}(\kappa _{l,m-1}\circ \lambda _{0,1})\right)
\end{split}
\]
and
\[
\begin{split}
& \left( \sum _{l,m=0}^{\infty }\Xi _{l,m}(\kappa _{l,m})\right)
  \Xi _{1,1}(\lambda _{1,1}) \\
& =\sum _{l,m=0}^{\infty }\sum _{k=0}^{m\wedge 1}k!
  \dbinom{m}{k}\dbinom{1}{k}
  \Xi _{l+1-k,m+1-k}\left( S_{m-k}^{l}\, _{1}^{1-k}
  (\kappa _{l,m}\circ _{k}\lambda _{1,1})\right) \\
& =\sum _{m=1}^{\infty }\Xi _{0,m}\left( mS_{m-1}^{0}\, ^{0}_{1}
  (\kappa _{0,m}\circ _{1}\lambda _{1,1})\right) \\
& \quad+\sum _{l,m=1}^{\infty }\Xi _{l,m}\left( S_{m-1}^{l-1}\, _{1}^{1}
  (\kappa _{l-1,m-1}\circ \lambda _{1,1})+mS_{m-1}^{l}\, _{1}^{0}
  (\kappa _{l,m}\circ _{1}\lambda _{1,1})\right) . 
\end{split}
\]

Thus,
\begin{equation}
\begin{split}
& \left( \sum _{l,m=0}^{\infty }\Xi _{l,m}(\kappa _{l,m})\right)\pi (\Psi ) \\
& =\Xi _{0,0}(\kappa _{0,1}\circ _{1}\lambda _{1,0}) \\
& \quad +\sum _{l=1}^{\infty }\Xi _{l,0}\left( S_{0}^{l-1}\, _{0}^{1}
    (\kappa _{l-1,0}\circ \lambda _{1,0})+S_{0}^{l}\, _{0}^{0}
    (\kappa _{l,1}\circ _{1}\lambda _{1,0})\right) \\
& \quad +\sum _{m=1}^{\infty }\Xi _{0,m} \Big( m\, S_{m-1}^{0}\, ^{0}_{1}
    (\kappa _{0,m}\circ _{1}\lambda _{1,1}) \\
& \qquad +(m+1)\, S_{m}^{0}\, _{0}^{0}
    (\kappa _{0,m+1}\circ _{1}\lambda _{1,0})-S_{m-1}^{0}\, _{1}^{0}
    (\kappa _{0,m-1}\circ \lambda _{0,1}) \Big)  \\
& \quad +\sum _{l,m=1}^{\infty }\Xi _{l,m} \Big( S_{m-1}^{l-1}\, _{1}^{1}
    (\kappa _{l-1,m-1}\circ \lambda _{1,1})+mS_{m-1}^{l}\, _{1}^{0}
    (\kappa _{l,m}\circ _{1}\lambda _{1,1}) \\
& \qquad +S_{m}^{l-1}\, _{0}^{1}(\kappa _{l-1,m}\circ \lambda _{1,0})
  +(m+1)S_{m}^{l}\, _{0}^{0}(\kappa _{l,m+1}\circ _{1}\lambda _{1,0})
  -S_{m-1}^{l}\, _{1}^{0}(\kappa _{l,m-1}\circ \lambda _{0,1}) \Big)
\label{eq:Theta_Psi} 
\end{split} 
\end{equation}
On the other hand, it follows that 
\begin{equation}
\widetilde{\pi }(\Psi )\left( \sum _{l,m=0}^{\infty }\Xi _{l,m}(\kappa _{l,m})\right) =-\left\{ \left( \sum _{l,m=0}^{\infty }\Xi _{l,m}(\kappa _{l,m})^{*}\right) \pi (\Psi )\right\} ^{*}
\label{eq:Psi_Theta_equal_(Theta_Psi)_star}
\end{equation}
as a continuous linear operator from \( (E) \) to \( (E)^{*} \). When we consider
\[
\begin{split}
s_{m+m'-k,l+l'-k}&t_{m+m'-k,l+l'-k}(S^{l}_{m-k}\, _{m'}^{l'-k}(\kappa \circ _{k}\lambda ))\\
  &=s_{m+m'-k,l+l'-k}(S_{l'-k}^{m'}\, _{l}^{m-k}(t_{m',l'}(\lambda )\circ _{k}t_{m,l}(\kappa )))
\end{split}
\]
for all \( \kappa \in \left( E^{\otimes (l+m)}\right) ^{*} \) and 
\( \lambda \in E^{\otimes l'}\otimes \left( E^{\otimes m'}\right) ^{*} \), 
the relations \eqref{eq:Theta_Psi}, 
\eqref{eq:Psi_Theta_equal_(Theta_Psi)_star},
and proposition \ref{thm:adjoint_of_integral_kernel_op} imply
\[
\begin{split}
&\widetilde{\pi }(\Psi ) \left( \sum _{l,m=0}^{\infty }\Xi _{l,m}(\kappa _{l,m})\right) \\
& =-\Xi _{0,0}(\lambda _{0,1}\circ _{1}\kappa _{1,0}) \\
& \quad +\sum _{l=1}^{\infty }\Xi _{l,0}
   \Big( l\, S_{0}^{1}\, _{0}^{l-1}(\lambda _{1,1}\circ _{1}\kappa _{l,0})
   +S_{0}^{1}\, _{0}^{l-1}(\lambda _{1,0}\circ \kappa _{l-1,0})
   -(l+1)S_{0}^{0}\, _{0}^{l}(\lambda _{0,1}\circ _{1}\kappa _{l+1,0})\Big) \\
& \quad +\sum _{m=1}^{\infty }\Xi _{0,m}
   \Big( -S_{1}^{0}\, _{m-1}^{0}(\lambda _{0,1}\circ \kappa _{0,m-1})
   -S_{0}^{0}\, _{m}^{0}(\lambda _{0,1}\circ _{1}\kappa _{1,m}) \Big) \\
& \quad +\sum _{l,m=1}^{\infty }\Xi _{l,m}
   \Big( S^{1}_{1}\, _{m-1}^{l-1}(\lambda _{1,1}\circ \kappa _{l-1,m-1})
   +l\, S_{0}^{1}\, _{m}^{l-1}(\lambda _{1,1}\circ _{1}\kappa _{l,m}) \\
& \qquad +S_{0}^{1}\, _{m}^{l-1}(\lambda _{1,0}\circ \kappa _{l-1,m})
   -S_{1}^{0}\, _{m-1}^{l}(\lambda _{0,1}\circ \kappa _{l,m-1})
   -(l+1)S_{0}^{0}\, _{m}^{l}(\lambda _{0,1}\circ _{1}\kappa _{l+1,m}) \Big) .
\end{split}
\]
Therefore we have 
\begin{gather*}
\begin{split}
s_{l,0}&\left( S_{0}^{l-1}\, _{0}^{1}(\kappa _{l-1,0}\circ \lambda _{1,0})
   +S_{0}^{l}\, _{0}^{0}(\kappa _{l,1}\circ _{1}\lambda _{1,0})\right) \\
& =s_{l,0}\left( l\, S_{0}^{1}\, _{0}^{l-1}(\lambda _{1,1}\circ _{1}\kappa _{l,0})
  +S_{0}^{1}\, _{0}^{l-1}(\lambda _{1,0}\circ \kappa _{l-1,0})
  -(l+1)S_{0}^{0}\, _{0}^{l}(\lambda _{0,1}\circ _{1}\kappa _{l+1,0})\right) ,
\end{split} \\
\begin{split}
s_{0,m}&\Big( m\, S_{m-1}^{0}\, ^{0}_{1}(\kappa _{0,m}\circ _{1}\lambda _{1,1})
   +(m+1)\, S_{m}^{0}\, _{0}^{0}(\kappa _{0,m+1}\circ _{1}\lambda _{1,0})
   -S_{m-1}^{0}\, _{1}^{0}(\kappa _{0,m-1}\circ \lambda _{0,1})\Big) \\
&=s_{0,m}\Big( -S_{1}^{0}\, _{m-1}^{0}(\lambda _{0,1}\circ \kappa _{0,m-1})
   -S_{0}^{0}\, _{m}^{0}(\lambda _{0,1}\circ _{1}\kappa _{1,m})\Big) ,
\end{split} \\
\begin{split}
s_{l,m}&\Big( S_{m-1}^{l-1}\, _{1}^{1}(\kappa _{l-1,m-1}\circ \lambda _{1,1})
    +mS_{m-1}^{l}\, _{1}^{0}(\kappa _{l,m}\circ _{1}\lambda _{1,1}) \\
& \quad +S_{m}^{l-1}\, _{0}^{1}(\kappa _{l-1,m}\circ \lambda _{1,0})
  +(m+1)S_{m}^{l}\, _{0}^{0}(\kappa _{l,m+1}\circ _{1}\lambda _{1,0})
  -S_{m-1}^{l}\, _{1}^{0}(\kappa _{l,m-1}\circ \lambda _{0,1}) \Big) \\
& =s_{l,m} \Big( S^{1}_{1}\, _{m-1}^{l-1}(\lambda _{1,1}\circ \kappa _{l-1,m-1})
  +l\, S_{0}^{1}\, _{m}^{l-1}(\lambda _{1,1}\circ _{1}\kappa _{l,m}) \\
& \quad +S_{0}^{1}\, _{m}^{l-1}(\lambda _{1,0}\circ \kappa _{l-1,m})
  -S_{1}^{0}\, _{m-1}^{l}(\lambda _{0,1}\circ \kappa _{l,m-1})
  -(l+1)S_{0}^{0}\, _{m}^{l}(\lambda _{0,1}\circ _{1}\kappa _{l+1,m})\Big) ,
\end{split}
\end{gather*}
and
\begin{equation*}
-\lambda _{0,1}\circ _{1}\kappa _{1,0}=\kappa _{0,1}\circ _{1}\lambda _{1,0}
\end{equation*}
for all \( l,m\geq 1 \). On the other hand, from \( \lambda _{1,0} \), \( \lambda _{0,1}\in E \)
and the definition of \( S^{\alpha }_{\beta }\, ^{\alpha '}_{\beta '} \) ,
we have
\begin{gather*}
s_{l,0}\left( S_{0}^{1}\, ^{l-1}_{0}(\kappa _{l-1,0}\circ \lambda _{1,0})\right) 
  =s_{l,0}\left( S_{0}^{l-1}\, ^{1}_{0}(\lambda _{1,0}\circ \kappa _{l-1,0})\right) ,\\
s_{0,m}\Big( S_{1}^{0}\, ^{0}_{m-1}(\lambda _{0,1}\circ \kappa _{0,m-1})\Big) 
  =s_{0,m}\Big( S_{m-1}^{0}\, ^{0}_{1}(\kappa _{0,m-1}\circ \lambda _{0,1})\Big) ,\\
s_{l,m}\left( S_{0}^{1}\, _{m}^{l-1}(\lambda _{1,0}\circ \kappa _{l-1,m})\right) 
  =s_{l,m}\left( S_{m}^{l-1}\, _{0}^{1}(\kappa _{l-1,m}\circ \lambda _{1,0})\right) ,\\
s_{l,m}\left( S_{1}^{0}\, ^{l}_{m-1}(\lambda _{0,1}\circ \kappa _{l,m-1})\right) 
  =s_{l,m}\left( S_{m-1}^{l}\, ^{0}_{1}(\kappa _{l,m-1}\circ \lambda _{0,1})\right) ,\quad and\\
s_{l,m}\left( S_{1}^{1}\, ^{l-1}_{m-1}(\lambda _{1,1}\circ \kappa _{l-1,m-1})\right) 
  =s_{l,m}\left( S_{m-1}^{l-1}\, ^{1}_{1}(\kappa _{l-1,m-1}\circ \lambda _{1,1}) \right)
\end{gather*}
for all \( l,m\geq 1 \). 
Hence we obtain \eqref{eq:irr-iff-condition(2)} \( - \) \eqref{eq:irr-iff-condition(1)}.
\end{proof}

\begin{cor}
\( \kappa _{l,m} \) , \( l,m \geq 1\) satisfy the following conditions:
\begin{gather}
s_{l,0}\left( S_{0}^{1}\, _{0}^{l-1}(\lambda _{1,1}\circ _{1}\kappa _{l,0})\right) =0,
\label{eq:irr-iff-condition(2)-cor} \\
s_{0,m}\Big( S_{m-1}^{0}\, _{1}^{0}(\kappa _{0,m}\circ _{1}\lambda _{1,1})\Big) =0,\quad and
\label{eq:irr-iff-condition(3)-cor} \\
l\, s_{l,m}\left( S_{0}^{1}\, _{m}^{l-1}(\lambda _{1,1}\circ _{1}\kappa _{l,m})\right) =m\, s_{l,m}\left( S_{m-1}^{l}\, _{1}^{0}(\kappa _{l,m}\circ _{1}\lambda _{1,1})\right) .
\label{eq:irr-iff-condition(4)-cor}
\end{gather}
\end{cor}
\begin{proof}
\( C^{\infty }(M,\mathfrak {g}) \)
contains constant functions. Thus if \( \Psi (x)=\mathrm{Const}.\in \mathfrak {g} \)
i.e., \( \lambda _{1,0}=\lambda _{0,1}=0 \), then we obtain 
\eqref{eq:irr-iff-condition(2)-cor} \( - \) \eqref{eq:irr-iff-condition(4)-cor} .
\end{proof}

Now, for \( l\geq 0 \), \( m\geq 1 \), we have
\[
\begin{split}
S&_{m-1}^{l}\, _{1}^{0}(\kappa _{l,m}\circ _{1}\lambda _{1,1}) \\
&= \sum _{\mathbf{i},\mathbf{j},\mathbf{j}'}\sum _{\mathbf{h}}
   \left\langle \kappa _{l,m},\, e(\mathbf{i})\otimes e(\mathbf{j})
   \otimes e(\mathbf{h})\right\rangle \left\langle e(\mathbf{h}),\, 
   V(\Psi )e(\mathbf{j}')\right\rangle 
   e(\mathbf{i})\otimes e(\mathbf{j})\otimes e(\mathbf{j}') \\
& =\sum _{\mathbf{i},\mathbf{j},\mathbf{j}'}\left\langle \kappa _{l,m},\, e(\mathbf{i})
   \otimes e(\mathbf{j})\otimes V(\Psi )e(\mathbf{j}')\right\rangle 
   e(\mathbf{i})\otimes e(\mathbf{j})\otimes e(\mathbf{j}') \\
& =(\mathrm{id}_{l+m-1}\otimes V(\Psi ))^{*}\kappa _{l,m}
  =(\mathrm{id}_{l+m-1}\otimes V(\Psi )^{*})\kappa _{l,m}. 
\end{split}
\]
In the same manner, it follows that 
\[
S_{0}^{1}\, _{m}^{l-1}(\lambda _{1,1}\circ \kappa _{l,m})=-(V(\Psi )\otimes \mathrm{id}_{l+m-1})^{*}\kappa _{l,m}=-(V(\Psi )^{*}\otimes \mathrm{id}_{l+m-1})\kappa _{l,m}\]
for \( l\geq 1 \), \( m\geq 0 \). 

On the other hand, we have
\begin{equation}
\begin{split}
s_{n}(A\otimes \mathrm{id}_{n-1})s_{n}&
   =s_{n}(\mathrm{id}\otimes A\otimes \mathrm{id}_{n-2})s_{n}
   =\ldots =s_{n}(\mathrm{id}_{n-1}\otimes A)s_{n}\\
  &=\frac{1}{n}d\Gamma _{\mathrm{b}}(A)^{(n)}.
    \label{eq:symmetrizer_and_second_quantization} 
\end{split}
\end{equation}
for \( A\in \mathcal{L} (E^{*},E^{*}) \) by direct computation. 
From \eqref{eq:symmetrizer_and_second_quantization}, 
\eqref{eq:irr-iff-condition(2)}\( - \)\eqref{eq:irr-iff-condition(4)}
are equivalent to 
\begin{equation}
d\Gamma _{\mathrm{b}}(V(\Psi )^{*})^{(l)}\kappa _{l,0}
  =s_{l,0}\Big( (l+1)S_{0}^{0}\, _{0}^{l}(\lambda _{0,1}\circ _{1}\kappa _{l+1,0})
   +S_{0}^{l}\, _{0}^{0}(\kappa _{l,1}\circ \lambda _{1,0})\Big) ,
\label{eq:irr-iff-condition(2)-ver2}
\end{equation}
\begin{equation}
\begin{split}
-d\Gamma _{\mathrm{b}}&(V(\Psi )^{*})^{(m)}\kappa _{0,m} \\
   &=s_{0,m}\Big( (m+1)S^{0}_{m}\, _{0}^{0}(\kappa _{0,m+1}\circ _{1}\lambda _{1,0})
    +S^{0}_{0}\, ^{0}_{m}(\lambda _{0,1}\circ _{1}\kappa _{1,m})\Big) ,\quad and
\label{eq:irr-iff-condition(3)-ver2}
\end{split}
\end{equation}
\begin{equation}
\begin{split}
\Big( & d\Gamma _{\mathrm{b}}(V(\Psi )^{*})^{(l)}\otimes \mathrm{id}_{m}
  +\mathrm{id}_{l}\otimes d\Gamma _{\mathrm{b}}(V(\Psi )^{*})^{(m)}\Big) \kappa _{l,m} \\
& =s_{l,m}\left( (l+1)S_{0}^{0}\, _{m}^{l}(\lambda _{0,1}\circ _{1}\kappa _{l+1,m})
  +(m+1)S_{m}^{l}\, _{0}^{0}(\kappa _{l,m+1}\circ _{1}\lambda _{1,0})\right) 
\label{eq:irr-iff-condition(4)-ver2}
\end{split}
\end{equation}
respectively for all \( \Psi \in C^{\infty }(M,\mathfrak {g}) \). 
In particular,
\eqref{eq:irr-iff-condition(2)-cor}\( - \)\eqref{eq:irr-iff-condition(4)-cor} become
\begin{equation}
\label{eq:irr-iff-condition(2)-cor-ver2}
d\Gamma _{\mathrm{b}}(V(\Psi )^{*})^{(l)}\kappa _{l,0}=0,
\end{equation}
\begin{equation}
\label{eq:irr-iff-condition(3)-cor-ver2}
d\Gamma _{\mathrm{b}}(V(\Psi )^{*})^{(m)}\kappa _{0,m}=0,
\end{equation}
\begin{equation}
\label{eq:irr-iff-condition(4)-cor-ver2}
\left\{ (d\Gamma _{\mathrm{b}}(V(\Psi )^{*})^{(l)}\otimes \mathrm{id}_{m}+\mathrm{id}_{l}\otimes d\Gamma _{\mathrm{b}}(V(\Psi )^{*})^{(m)}\right\} \kappa _{l,m}=0.
\end{equation}

\begin{lem}
For \( l,m\geq 0 \), \( l+m\neq 0 \), we have
\begin{equation}
\kappa _{l,m}=\sum \left\langle \kappa _{l,m},
        \widehat{e}(\mathbf{i},\mathbf{j})\otimes \widehat{e}(\mathbf{i}',\mathbf{j}')
       \right\rangle \widehat{e}(\mathbf{i},\mathbf{j})\otimes
       \widehat{e}(\mathbf{i}',\mathbf{j}'),
\label{eq:series_of_kappa_(l,m)}
\end{equation}
where the sum is over all \( (\mathbf{i},\mathbf{j})\in \Lambda (l) \) and
\( (\mathbf{i}',\mathbf{j}')\in \Lambda (m) \) satisfying the condition:
\[
   \sum _{1\leq p\leq N_{2}}\alpha _{p}(H)
   (n_{p,+}(\mathbf{j})-n_{p,-}(\mathbf{j})
   +n_{p,+}(\mathbf{j}')-n_{p,-}(\mathbf{j}'))=0
\]
for all \( H\in \mathfrak {h} \). Here
\begin{gather*}
 n_{p,+}(\mathbf{j})
    :=\#\{q\in \{1,2,\ldots ,n\}\, |\, j_{q}=N_{1}+p\}, \\
 n_{p,-}(\mathbf{j})
    :=\#\{q\in \{1,2,\ldots ,n\}\, |\, j_{q}=N_{1}+N_{2}+p\}. 
\end{gather*}
for \( \mathbf{j}=(j_{1},\ldots ,j_{n})\in \{1,2,\ldots ,N\}^{n} \)
and \( 1\leq p\leq N_{2} \) .
\end{lem}

\begin{proof}
Since
\[
d\Gamma _{\mathrm{b}}(V(H)^{(n)})\widehat{e}(\mathbf{i},\mathbf{j})
  =\sum _{1\leq p\leq N_{2}}\alpha _{p}(H)(n_{p,+}(\mathbf{j})
   -n_{p,-}(\mathbf{j}))\widehat{e}(\mathbf{i},\mathbf{j})\]
for \( H\in \mathfrak {h} \), we have
\begin{equation}
\begin{split}
&\big\{ d\Gamma _{\mathrm{b}}(V(H)^{*})^{(l)}\otimes
   \mathrm{id}_{m} \big\} \kappa _{l,m} \\
&= \sum _{(\mathbf{i},\mathbf{j})\in \Lambda (l)}
   \sum _{(\mathbf{i}',\mathbf{j}')\in \Lambda (m)} \sum _{1\leq p\leq N_{2}}
   \Big\langle
     \kappa _{l,m}, \big\{ d\Gamma _{\mathrm{b}}(V(H))^{(l)}
     \otimes \mathrm{id}_{m} \big\} 
     \big\{ \widehat{e}(\mathbf{i},\mathbf{j}) \otimes \widehat{e}(\mathbf{i}',\mathbf{j}') \big\}
   \Big\rangle
   \widehat{e}(\mathbf{i},\mathbf{j})\otimes \widehat{e}(\mathbf{i}',\mathbf{j}') \\
&= \sum _{(\mathbf{i},\mathbf{j})}\sum _{(\mathbf{i}',\mathbf{j}')\in \Lambda (m)}
   \sum _{1\leq p\leq N_{2}} \alpha _{p}(H)(n_{p,+}(\mathbf{j})-n_{p,-}(\mathbf{j}))
   \left\langle \kappa _{l,m},\widehat{e}(\mathbf{i},\mathbf{j})
    \otimes \widehat{e}(\mathbf{i}',\mathbf{j}')\right\rangle
   \widehat{e}(\mathbf{i},\mathbf{j})\otimes \widehat{e}(\mathbf{i}',\mathbf{j}')
\label{eq:second_q_op_act_kappa(1)} 
\end{split}
\end{equation}
where \( (\mathbf{i},\mathbf{j}) \) runs over the whole of \( \Lambda (l) \) 
satisfying the conditions: there exists \( p \) such that \( n_{p,+}(\mathbf{j})\neq n_{p,-}(\mathbf{j}) \).
In the same manner, 
\begin{equation}
\begin{split}
& \left(\mathrm{id}_{l}\otimes d\Gamma _{\mathrm{b}}(V(H)^{*})^{(m)}\right)
   \kappa _{l,m} \\
& =\sum _{(\mathbf{i},\mathbf{j})\in \Lambda (l)}\sum _{(\mathbf{i}',\mathbf{j}')}
   \sum _{1\leq p\leq N_{2}}
   \alpha _{p}(H)(n_{p,+}(\mathbf{j}')-n_{p,-}(\mathbf{j}'))
   \left\langle \kappa _{l,m},\widehat{e}(\mathbf{i},\mathbf{j})
     \otimes \widehat{e}(\mathbf{i}',\mathbf{j}')\right\rangle 
   \widehat{e}(\mathbf{i},\mathbf{j})\otimes \widehat{e}(\mathbf{i}',\mathbf{j}').
\label{eq:second_q_op_act_kappa(2)}
\end{split}
\end{equation}
where \( (\mathbf{i}',\mathbf{j}') \) runs over the whole of \( \Lambda (m) \) satisfying
the conditions: there exists \( p \) such that \( n_{p,+}(\mathbf{j}')\neq n_{p,-}(\mathbf{j}') \). Thus we obtain \eqref{eq:series_of_kappa_(l,m)} 
from \eqref{eq:irr-iff-condition(4)-cor-ver2}. 
\end{proof}

When the operator \( d\Gamma _{\mathrm{b}}(V(u_{k+N_{1}})^{*}) \), 
\( 1\leq k\leq N_{2} \) act \( \kappa _{l,m} \), we need the following lemma.

\begin{lem}
\[
V(X_{\alpha})^{*}|E=-V(X_{-\alpha}),
\]
that is, if \( 1\leq k\leq N_{2} \), then
\[
V(u_{k+N_{1}})^{*}|E=-V(u_{k+N_{1}+N_{2}}).
\]
\end{lem}

\begin{proof}
Let \( U_{\alpha } \) and \(V_{\alpha } \), \( \alpha \in \Delta '\) be elements defined by
\begin{equation}
  X_{\alpha }=\frac{U_{\alpha }+\sqrt{-1}V_{\alpha }}{|U_{\alpha }+\sqrt{-1}V_{\alpha }|_{\mathfrak {g}}},
  \quad X_{-\alpha }=\frac{U_{\alpha }-\sqrt{-1}V_{\alpha }}{|U_{\alpha }-\sqrt{-1}V_{\alpha }|_{\mathfrak {g}}} .
\label{eq:Rel_between_Weyl_b-and-Cartan-b}
\end{equation}
And let \( \mathfrak {g}_{\mathbf{R}} \) be a compact real form of \( \mathfrak {g}^{\mathrm{c}} \)
and \( \mathfrak {h}_{\mathbf{R}} \) be a Cartan subalgebra of \( \mathfrak {g}_{\mathbf{R}} \).
If \( \{ \sqrt{-1}H'_{i} \ | \ i=1,2,\ldots , N_{1} \} \) is a basis of
\( \mathfrak {h}_{\mathbf{R}} \), then
\[
  \{\sqrt{-1}H'_{1},\ldots ,\sqrt{-1}H'_{N_{1}},U_{\alpha },V_{\alpha }\, 
  |\, \sqrt{-1}H'_{i}\in \mathfrak {h}_{\mathbf{R}},\, \alpha \in \Delta '\}
\]
is a  basis of \( \mathfrak {g}_{\mathbf{R}} \), called Cartan-Weyl basis. 
(More details of this basis are in \cite{Samelson}.)

Since \( \mathfrak {g} \) is compact, there exists an automorphism \( \sigma  \)
of \( \mathfrak {g}^{\mathbf{c}} \) such that \( \sigma (\mathfrak {g}_{\mathbf{R}})=\mathfrak {g} \).
Hence we can regard the Cartan-Weyl basis of \( \mathfrak {g}_{\mathbf{R}} \)
as a basis of \( \mathfrak {g} \). 

From lemma \ref{thm:adjoint_op_of_ad(Z)}, for \( f \), \( g\in E \), we have
\[
\left\langle V(X_{\alpha })^{*}f,g\right\rangle =\left\langle \overline{f},V(X_{\alpha })g\right\rangle _{0}=\left\langle -V(\overline{X_{\alpha }})\overline{f},g\right\rangle _{0}=\left\langle -\overline{V(\overline{X_{\alpha }})\overline{f}},g\right\rangle .\]
If we note that the identification of \( E_{-p} \) and \( E^{*}_{p} \) under
the anti-linear isomorphism (see definition \ref{thm:property_of_self-adj_op_A}),
then this implies that \( V(X_{\alpha })^{*}f=-V(\overline{X_{\alpha }})f \).
Since \( \overline{X_{\alpha }}=X_{-\alpha } \) from \eqref{eq:Rel_between_Weyl_b-and-Cartan-b},
we obtain this lemma.
\end{proof}

\begin{lem}
\( \kappa _{1,0}=0 \) and \( \kappa _{0,1}=0 \).
\end{lem}
\begin{proof}
\eqref{eq:series_of_kappa_(l,m)} implies that
\[
\kappa _{1,0}=\sum _{(i,j)\in \Lambda (1);1\leq j\leq N_{1}}\left\langle \kappa _{1,0},e(i,j)\right\rangle e(i,j).
\]
For \( 1\leq k\leq N_{2} \) , if the operator
\( d\Gamma _{\mathrm{b}}(V(u_{k+N_{1}})^{*}) \)
act \( \kappa _{1,0} \), then we obtain 
\[
0=d\Gamma _{\mathrm{b}}(V(u_{k+N_{1}})^{*})\kappa _{1,0}=-\sum _{i=1}^{\infty }\sum _{1\leq j\leq N_{1}}\left\langle \kappa _{1,0},e(i,j)\right\rangle \alpha _{k}(u_{j})e(i,k+N_{1}+N_{2})
\]
with the help of
\[
\begin{split}
  V(u_{k+N_{1}})^{*}e(i,j)&=-V(u_{k+N_{1}+N_{2}})e(i,j)
   =-e_{i}\otimes \mathrm{ad}(u_{k+N_{1}+N_{2}})u_{j}\\
   &=-\alpha _{k}(u_{j})e(i,k+N_{1}+N_{2}).
\end{split}
\]
This implies
\[
  \sum _{1\leq j\leq N_{1}}\left\langle \kappa _{1,0},e(i,j)\right\rangle \alpha _{k}(u_{j})=0
\]
for all \( i\in \mathbf{N} \), \( k\in \{1,2,\ldots ,N_{2}\} \). Since \( \mathfrak {h}^{*} \)
is generated by linear combinations of \( \{\alpha _{k}\}_{k=1}^{N_{2}} \),
we can choose basis \( \{\alpha _{k_{1}},\ldots ,\alpha _{k_{N_{1}}}\} \) of
\( \mathfrak {h}^{*} \). Then the matrix \( (\alpha _{k_{i}}(u_{j}))_{1\leq i,j\leq N_{1}}\in \mathrm{Mat}(N_{1},\mathbf{C}) \)
is invertible. Therefore
\[
\left\langle \kappa _{1,0},e(i,j)\right\rangle =0\]
for all \( i\in \mathbf{N} \) and \( j\in \{1,2,\ldots ,N_{1}\} \), i.e., \( \kappa _{1,0}=0 \).
In the same manner, \( \kappa _{0,1}=0 \).
\end{proof}

\begin{lem}
\( \kappa _{l,1}=0 \) and \( \kappa _{1,m}=0 \).
\end{lem}
\begin{proof}
In \eqref{eq:irr-iff-condition(4)-cor-ver2}, let \( \Psi =u_{k+N_{1}} \) for
\( 1\leq k\leq N_{2} \) and \( m=1 \). When we consider \eqref{eq:series_of_kappa_(l,m)},
we have
\[
\begin{split}
& (d\Gamma _{\mathrm{b}}(V(u_{k+N_{1}})^{*})^{(l)}\otimes \mathrm{id})\kappa _{l,1} \\
&= \sum _{(\mathbf{i},\mathbf{j})}\, \sum _{i\in \mathbf{N}}\, \sum _{1\leq j\leq N_{1}}
   \left\langle \kappa _{l,1},\widehat{e}(\mathbf{i},\mathbf{j})\otimes e(i,j)\right\rangle
   [d\Gamma _{\mathrm{b}}(V(u_{k+N_{1}})^{*})^{(l)}
   \widehat{e}(\mathbf{i},\mathbf{j})]\otimes e(i,j),
\end{split}
\]
and
\[
\begin{split}
& (\mathrm{id}_{l}\otimes d\Gamma _{\mathrm{b}}(V(u_{k+N_{1}})^{*})^{(1)})\kappa _{l,1} \\
&= \sum _{(\mathbf{i},\mathbf{j})}\, \sum _{i\in \mathbf{N}}\, \sum _{1\leq j\leq N_{1}}
   \left\langle \kappa _{l,1},\widehat{e}(\mathbf{i},\mathbf{j})\otimes e(i,j)\right\rangle
   \alpha _{k}(u_{j})\widehat{e}(\mathbf{i},\mathbf{j})\otimes e(i,k+N_{1}+N_{2}),
\end{split}
\]
where \( (\mathbf{i},\mathbf{j}) \) runs over the whole of \( \Lambda (l) \) satisfying
\begin{equation}
\sum _{1\leq p\leq N_{2}}\alpha _{p}(H)(n_{p,+}(\mathbf{j})-n_{p,-}(\mathbf{j}))=0.
\label{eq:condition_(i,j)}
\end{equation}
If \( 1\leq j\leq N_{1} \) and \( 1\leq k\leq N_{2} \), then \( [d\Gamma _{\mathrm{b}}(V(u_{k+N_{1}})^{*})^{(l)}\widehat{e}(\mathbf{i},\mathbf{j})]\otimes e(i,j) \)
and \( \widehat{e}(\mathbf{i},\mathbf{j})\otimes e(i,k+N_{1}+N_{2}) \) are orthogonal
each other with respect to the inner product on
\( H(M,\mathfrak {g})^{\widehat{\otimes }l} \otimes H(M,\mathfrak {g}) \).
Thus, \( (\mathrm{id}_{l}\otimes d\Gamma _{\mathrm{b}}(V(u_{k+N_{1}})^{*})^{(1)})\kappa _{l,1}=0 \)
and hence
\begin{equation}
 \sum _{1\leq j\leq N_{1}}\left\langle \kappa _{l,1},
 \widehat{e}(\mathbf{i},\mathbf{j})\otimes e(i,j)\right\rangle \alpha _{k}(u_{j})=0
\label{eq:kappa_(l,1)-condition(1)}
\end{equation}
for all \( i\in \mathbf{N} \), \( 1\leq k\leq N_{2} \), 
and all \( (\mathbf{i},\mathbf{j})\in \Lambda (l) \)
satisfying \eqref{eq:condition_(i,j)}. 

Since we can select \( k_{1} \), \( k_{2} \),\( \ldots  \), \( k_{N_{1}}\in \{1,2,\ldots ,N_{2}\} \)
such that a matrix \( (\alpha _{k_{i}}(u_{j}))_{1\leq i,j\leq N_{1}}\in \mathrm{Mat}(N_{1},\mathbf{C}) \)
is invertible, we obtain
\[
\left\langle \kappa _{l,1},\widehat{e}(\mathbf{i},\mathbf{j})\otimes e(i,j)\right\rangle =0\]
for all \( i\in \mathbf{N} \), and \( 1\leq j\leq N_{1} \) and all \( (\mathbf{i},\mathbf{j})\in \Lambda (l) \)
satisfying \eqref{eq:condition_(i,j)}, i.e., \( \kappa _{l,1}=0 \). 

In the same manner, we also have \( \kappa _{1,m}=0 \).
\end{proof}

\begin{lem}
\( \kappa _{l,0}=0 \) and \( \kappa _{0,m}=0 \) for all \( l,m\geq 1 \).
\end{lem}

\begin{proof}
We prove them by the induction. We have only to show \( \kappa _{l,0}=0 \).
The case of \( \kappa _{1,0}=0 \) has been proved. Let \( \kappa _{l,0}=0 \).
Since \( \kappa _{l,1}=0 \) and \eqref{eq:irr-iff-condition(4)-ver2}, we have
\[
\begin{split}
0 & = s_{l,0}\left( (l+1)S_{0}^{0}\, _{0}^{l}(\lambda _{0,1}\circ _{1}\kappa _{l+1,0})\right) \\
  & = \sum _{(\mathbf{i},\mathbf{j})\in \Lambda (l)}
      \sum _{(i,j)\in \Lambda (1)}\left\langle \lambda _{0,1},e(i,j)
      \right\rangle \left\langle \kappa _{l+1,0},e(i,j)\otimes 
      \widehat{e}(\mathbf{i},\mathbf{j})\right\rangle \widehat{e}(\mathbf{i},\mathbf{j})\\
 & = \sum _{(\mathbf{i},\mathbf{j})\in \Lambda (l)}\left\langle 
     \kappa _{l+1,0},\lambda _{0,1}\otimes \widehat{e}(\mathbf{i},\mathbf{j})
     \right\rangle \widehat{e}(\mathbf{i},\mathbf{j})
\end{split}
\]
and hence
\begin{equation}
\left\langle \kappa _{l+1,0},\lambda _{0,1}\otimes \widehat{e}(\mathbf{i},\mathbf{j})\right\rangle =0
\label{eq:kappa_(l+1,0)--lambda(0,1)}
\end{equation}
for all \( (\mathbf{i},\mathbf{j})\in \Lambda (l) \). This implies that 
\begin{gather}
\left\langle \kappa _{l+1,0},d\Psi \otimes \widehat{e}(\mathbf{i},\mathbf{j})\right\rangle =0,
   \label{eq:kappa_and_d(Psi)} \\
\left\langle \kappa _{l+1,0},V(\Psi )d\Psi '\otimes \widehat{e}(\mathbf{i},\mathbf{j})\right\rangle =0 
   \label{eq:kappa_and_V(Psi)dPsi'}
\end{gather}
for all \( (\mathbf{i},\mathbf{j})\in \Lambda (l) \), and \( \Psi  \), 
\( \Psi '\in C^{\infty }(M,\mathfrak {g}) \).
\eqref{eq:kappa_and_d(Psi)} is obvious. We show \eqref{eq:kappa_and_V(Psi)dPsi'}. 

For each \( \Psi  \), \( \Psi '\in C^{\infty }(M,\mathfrak {g}) \) and \( |s| \),\( |t|\ll 1 \),
there exists a unique \( \Phi _{s,t}\in C^{\infty }(M,\mathfrak {g}) \) such
that 
\[
\exp (t\Psi )\exp (s\Psi ')=\exp (\Phi _{s,t}).\]
Since
\[
\begin{split}
 d\Phi _{s,t}&=\beta (\exp (\Phi _{s,t}))=\beta (\exp (t\Psi )\exp (s\Psi '))\\
   &=V(\exp (t\Psi ))\beta (\exp (s\Psi '))+\beta (\exp (t\Psi )) \\
   &=sV(\exp (t\Psi ))d\Psi '+td\Psi
\end{split}
\]
and \eqref{eq:kappa_and_d(Psi)}, we have
\[
\begin{split}
0 & = \left\langle \kappa _{l+1,0},d\Phi _{s,t}\otimes \widehat{e}(\mathbf{i},\mathbf{j})\right\rangle \\
  & = s \left\langle \kappa _{l+1,0},V(\exp (t\Psi ))
      d\Psi '\otimes \widehat{e}(\mathbf{i},\mathbf{j})\right\rangle 
      +t\left\langle \kappa _{l+1,0},d\Psi \otimes \widehat{e}(\mathbf{i},\mathbf{j})\right\rangle \\
  & = s \left\langle \kappa _{l+1,0},V(\exp (t\Psi ))
      d\Psi '\otimes \widehat{e}(\mathbf{i},\mathbf{j})\right\rangle .
\end{split}
\]
Hence \( \kappa _{l+1,0} \) satisfies \eqref{eq:kappa_and_V(Psi)dPsi'} by considering
the differential of the above equation at \( t\in \mathbf{R} \). 

By the way, \( H(M,\mathfrak {g}) \) is generated by
\[
\{d\Psi ,\, V(\Psi )d\Psi '\, |\, \Psi ,\Psi '\in C^{\infty }(M,\mathfrak {g})\}.\]
(See lemma 3.5 of \cite{Albe-paper}.) Thus the following relation is a direct
consequence of \eqref{eq:kappa_and_d(Psi)} and \eqref{eq:kappa_and_V(Psi)dPsi'}.
\[
\left\langle \kappa _{l+1,0},e(i,j)\otimes \widehat{e}(\mathbf{i},\mathbf{j})\right\rangle =0\]
for all \( (i,j)\in \Lambda (1) \) and \( (\mathbf{i},\mathbf{j})\in \Lambda (l) \).
Therefore we obtain \( \kappa _{l+1,0}=0 \). 

In the same manner, we can show \( \kappa _{0,m}=0 \) for all \( m\geq 1 \).
\end{proof}

\begin{lem}
\( \kappa _{l,m}=0 \) for all \( (l,m)\in \mathbf{Z}_{\geq 0}^{2}\setminus \{(0,0)\} \), 
that is, we obtain theorem \ref{thm:main-theorem}
\end{lem}
\begin{proof}
We prove this statement by the induction. We have already shown the case of
\( l=1 \), i.e., \( \kappa _{1,m}=0 \) for all \( m\geq 0 \). Let \( \kappa _{l,m}=0 \)
for all \( m\geq 0 \). Then we show \( \kappa _{l+1,m}=0 \) for all \( m\geq 0 \).
Fix \( m\geq 0 \). Since \( \kappa _{l,m}=0 \) and \( \kappa _{l,m+1}=0 \)
and \eqref{eq:irr-iff-condition(4)}, we have
\[
\begin{split}
0 & = s_{l,m}\left( (l+1)S_{0}^{0}\, _{m}^{l}
      (\lambda _{0,1}\circ _{1}\kappa _{l+1,m})\right) \\
 & = (l+1)\sum _{(\mathbf{i},\mathbf{j})\in \Lambda (l)} 
          \sum_{(\mathbf{i}',\mathbf{j}')\in \Lambda (m)}
      \left\langle \kappa _{l+1,m},\lambda _{0,1}\widehat{\otimes }\, 
      \widehat{e}(\mathbf{i},\mathbf{j})
        \otimes \widehat{e}(\mathbf{i}',\mathbf{j}')\right\rangle
      \widehat{e}(\mathbf{i},\mathbf{j})\otimes \widehat{e}(\mathbf{i}',\mathbf{j}').
\end{split}
\]
This shows \( \left\langle \kappa _{l+1,m},\lambda _{0,1}
\widehat{\otimes }\, \widehat{e}(\mathbf{i},\mathbf{j})
\otimes \widehat{e}(\mathbf{i}',\mathbf{j}')\right\rangle =0 \)
for all \( (\mathbf{i},\mathbf{j})\in \Lambda (l),\, (\mathbf{i}',\mathbf{j}')\in \Lambda (m) \).
Thus we can show
\[
  \left\langle \kappa _{l+1,m},e(i,j)\, \widehat{\otimes} \, \widehat{e}
  (\mathbf{i},\mathbf{j})\otimes \widehat{e}(\mathbf{i}',\mathbf{j}')\right\rangle =0
\]
for all \( (i,j)\in \Lambda (1) \), \( (\mathbf{i},\mathbf{j})\in \Lambda (l) \),
and \( (\mathbf{i}',\mathbf{j}')\in \Lambda (m) \). This implies \( \kappa _{l+1,m}=0 \).
Since \( m\geq 0 \) is arbitrary, the proof has been completed.
\end{proof}

\begin{flushleft}
{\Large \bf Acknowledgements}
\end{flushleft}
I am grateful to Professor Taku Matsui for his many useful comments on this
paper. And I am grateful to Professors Un Cig Ji and Nobuaki Obata for discussing
the difference between the energy representation of \( C^{\infty }(S^{1},G) \)
in this paper and the factorial representation of the path group in \cite{Albe-paper}
and \cite{Albe_1-dim_case}.


\begin{thebibliography}{11}

\bibitem{Albe-NonCommDistTh}
S. Albeverio, R. H{\o}egh-Krohn, J. Marion, D. Testard
and B. Torr{\'e}sani: {\it Noncommutative distributions. Unitary
representation of gauge groups and algebras}. Marcel Dekker, Inc., 1993.

\bibitem{Albe-paper}
S. Albeverio, R. H{\o}egh-Krohn and D. Testard: {\it Irreducibility and reducibility
for the energy representation of the group of mappings of a Riemannian manifold
into a compact semisimple Lie group}. J. Funct. Anal. {\bf 41} (1981), no. 3,
378--396.

\bibitem{Albe_1-dim_case}
S. Albeverio, R. H{\o}egh-Krohn, D. Testard and A. Ver{\v s}hik: {\it Factorial
representations of path groups}. J. Funct. Anal. {\bf 51} (1983), no. 1, 115--131.
\bibitem{Bourbaki} N. Bourbaki:
{\it Lie groups and Lie algebras}. 2nd printing, English
translation. Springer-Verlag, 1989

\bibitem{Gelfand_et_al(1977)} I. Gelfand, M. Graev and A. Ver\v sik: 
{\it Representations of the group of smooth mappings of 
a manifold $X$ into a compact Lie group}. 
Compositio Math. {\bf 35} (1977), no. 3, 299--334.

\bibitem{Gelfand_et_al} I. Gelfand, M. Graev and A. Ver\v sik: 
{\it Representations of the group of functions taking values 
in a compact Lie group}. Compositio Math. {\bf 42}
(1980/81), no. 2, 217--243. 

\bibitem{Gilkey} P. Gilkey: {\it Invariance theory, the heat equation, 
and the Atiyah-Singer index theorem}. 
Second edition. CRC Press, 1995. 

\bibitem{Ismagilov-paper} R. Ismagilov: 
{\it Unitary representations of the group \( C^{\infty }_{0}(X,G) \),
\( G=SU_{2} \)}. Mat. Sb.{\bf 100} (1976), no. 1, 117--131;English transl.
in Math. USSR-Sb.{\bf 29}(1976), 105--117. 

\bibitem{Ismagilov-book} R. Ismagilov: 
{\it Representations of infinite-dimensional groups}. Translated
from the Russian manuscript by D. Deart. 
Translations of Mathematical Monographs, 152. 
American Mathematical Society, 1996. 

\bibitem{Ji-Obata} U.C. Ji and N. Obata: 
{\it Quantum white noise calculus}. Non-Commutativity,
Infinite-Dimensionality and Probability at the Crossroads 
(N. Obata, T. Matsui and A. Hora, Eds.), pp.143-191, World Scientific, 2002.

\bibitem{Kuo}H.-H. Kuo: {\it White noise distribution theory}. CRC Press, 1996.

\bibitem{Obata}N. Obata: {\it White noise calculus and Fock space}. 
Lecture Notes in Math, 1577. Springer-Verlag, 1994. 

\bibitem{Salomonsen} G. Salomonsen: {\it Equivalence of Sobolev spaces}. 
Results Math. {\bf 39} (2001), no. 1-2, 115--130.

\bibitem{Samelson} H. Samelson: {\it Notes on Lie algebras}. 
Second edition. Springer-Verlag, 1990.
 
\bibitem{Versik_et_al}
A.M.Ver\v sik, I.M.Gelfand, and M.I.Graev : 
{\it Representations of the group ${\rm SL}(2,\,R)$, 
where $R$ is a ring of functions}. (Russian) 
Uspehi Mat. Nauk {\bf 28} (1973), no. 5(173), 83--128. 

\bibitem{Wassermann} A. Wassermann: 
{\it Operator algebras and conformal field theory. III. Fusion
of positive energy representations of {\rm LSU(\( N\))} using
bounded operators}. Invent. Math. {\bf133} (1998), no. 3, 467--538.

\end{thebibliography}
\end{document}